\documentclass[twocolumn,bibyear]{aa}

\usepackage[varg]{txfonts}

\newcommand{\nc}{\newcommand}

\nc{\Msun}{\ensuremath{\mathrm{M}_\odot}}
\nc{\Rsun}{\ensuremath{\mathrm{R}_\odot}}
\nc{\cmcub}{\mbox{cm$^{-3}$}}
\newcommand{\CII}{[C \,{\sc ii}]}

\nc\micron{\mbox{$\mu$m}}
\nc{\thCO}{$^{13}$CO}
\nc{\hcoplus}{HCO$^+$}
\newcommand{\OI}{[O \,{\sc i}]}

\nc{\cmsq}{\mbox{cm$^{-2}$}}
\nc{\kms}{\mbox{km~s$^{-1}$}}
\nc{\lsun}{\ensuremath{\mathrm{L}_\odot}}
\newcommand{\HII}{H \,{\sc ii}}
\nc{\CeiO}{C$^{18}$O}
\nc{\tex}{T$_{\rm ex}$}
\nc{\tkin}{T$_{\rm kin}$}
\nc{\rsun}{\ensuremath{\mathrm{R}_\odot}}

\newcommand{\cmark}{\color{black}}
\newcommand{\umark}{\color{black}}

\usepackage{graphicx}
\usepackage[varg]{txfonts}
\usepackage{natbib}
\usepackage[colorlinks=true,     linkcolor=blue, citecolor=blue, filecolor=blue, urlcolor=blue]{hyperref}

\begin{document} 

\title{SOFIA/upGREAT far-infrared spectroscopy of bright rimmed pillars in IC~1848}

\author{Dariusz C. Lis\inst{1}
  \and Rolf G\"{u}sten\inst{2}
  \and Paul F. Goldsmith\inst{1}
  \and Yoko Okada\inst{3}
  \and Youngmin Seo\inst{1}
  \and Helmut Wiesemeyer\inst{2}
  \and Marc Mertens\inst{2,3}
  }

\institute{Jet Propulsion Laboratory, California Institute of Technology, 4800 Oak Drove Drive, Pasadena, CA 91109, USA
  \and Max-Planck-Institut für Radioastronomie, Auf dem H\"{u}gel 69, D-53121 Bonn, Germany
  \and I. Physikalisches Institut, Universit\"{a}t zu K\"{o}ln, Z\"{u}lpicher Stra{\ss}e 77, 50937 K\"{o}ln, Germany
}

\date{Received 1 August 2024 / Accepted 25 September 2024 / $\copyright 2024$ All rights reserved.}

\abstract{
Using the upGREAT instrument on SOFIA, we have imaged the \CII\ 158~$\mu$m fine structure line emission in bright-rimmed pillars located at the southern edge of the IC1848 \HII\ region, and carried out pointed observations of the \OI\ 63 and 145~$\mu$m fine structure lines toward selected positions. The observations are used to characterize the morphology, velocity field, and the physical conditions in the G1--G3 filaments. The velocity-resolved \CII\ spectra show evidence of a velocity shift at the head of the brightest G1 filament, possibly caused by radiation pressure from the impinging UV photons or the rocket effect of the evaporating gas. Archival \emph{Herschel} PACS and SPIRE data imply H$_2$ column densities in the range $10^{21} - 10^{22}$~cm$^{-2}$, corresponding to maximum visual extinction $A_V \simeq 10$~mag, and average H$_2$ volume density of $\simeq 4500$~cm$^{-3}$ in the filaments. The [C\,{\sc ii}] emission traces $\sim 17\%$ of the total H$_2$ column density, as derived from dust SED fits. PDR models are unable to explain the observed line intensities of the two \OI\ fine structure lines in IC1848, with the observed \OI\ 145~$\mu$m line being too strong compared to the model predictions. The \OI\ lines in IC1848 are overall weak and the signal-to-noise ratio is limited. However, our observations suggest that the \OI\ 63/145~$\mu$m intensity ratio is a sensitive probe of the physical conditions in photon dominated regions such as IC1848. These lines are thus excellent targets for future high-altitude balloon instruments, less affected by telluric absorption.
}
  

\titlerunning{Bright rimmed pillars in IC~1848}
\authorrunning{Lis et al.}

\maketitle 

\nolinenumbers

\section{Introduction}
\label{sec:intro}


The interaction of high-mass stars with their environment determines the rate of star formation and star formation efficiency of molecular clouds, and regulates the evolution of galaxies. Mechanical and radiative energy input from high-mass stars can stir up, ionize, and heat the gas, leading to the disruption of the natal molecular cloud within a few free-fall times (see \citealt{schneider20}, and references therein.) However, high-mass stars can also provide positive feedback, by compressing the gas and allowing gravity to overcome the turbulent and magnetic support that had prevented collapse and star formation. 

The aim of the SOFIA FEEDBACK legacy program \citep{schneider20, tiwari22, schneider23} was to quantify the effects of stellar feedback in a range of environments ranging from single OB stars, to small groups, to rich young clusters, and to mini-starburst complexes, by determining the physical conditions using observations of the far-infrared [C\,{\sc ii}] and [O\,{\sc i}] fine structure lines, combined with H$_2$ line observations from \emph{Spitzer}, and ground-based millimeter wave observations.

Photon-Dominated Regions (PDRs), the surface layers of molecular clouds exposed to strong fluxes of UV photons, dominate the IR spectra of star-forming clouds and galaxies, and provide a unique tool to study in detail the physical and chemical processes in diverse environments ranging from diffuse clouds, molecular cloud surfaces, globules, protoplanetary disks, planetary nebulae, and starburst galaxies. The JWST Early Release Science program PDRs4All \citep{berne22, peeters24, berne24} provides template datasets designed to identify key PDR characteristics in JWST spectra of star-forming regions in our Galaxy and beyond. As discussed in \cite{berne22}, PDRs are strong emitters in the near and mid-IR, where their key diagnostic signatures include (a) the Aromatic Infrared Bands (AIBs) attributed to PAHs, (b) continuum emission from very small carbonaceous grains (VSGs), (c) ro-vibrational and pure rotational lines of H$_2$, and (d) emission from atomic ions (e.g., S$^+$, Si$^+$, Fe$^+$).


Hydrodynamical numerical simulations \citep{mackey10} show that shadowing of ionizing radiation by an inhomogeneous density field can form so-called elephant trunks (ETs), or pillars of dense gas without the assistance of self-gravity or of ionization front and cooling instabilities. In some cases, the head of the ET is shown to recede from the radiation source more rapidly than the shadowed trunk which has not been exposed to radiation. For dynamically forming ETs, gas streams into the shadowed trunk past the head and is thus moving faster than, or at a comparable speed to, the pillar’s head. This could result in gas compression, triggering collapse and formation of next generation of stars. 
A combined Chandra/optical study aimed at quantifying triggered star formation in IC 1396A \citep{getman12} also argues for a radiation-driven implosion process persisting over several million years. These authors argue that the contribution of the triggered star formation for the entire HII region exceeds 14 – 25\% today and may have been higher over the lifetime of the HII region. Such triggering on the periphery of HII regions may thus be a significant mode of star formation in the Galaxy.

Magnetic fields have been suggested to have a significant influence on the development of pillars around H\,{\sc ii} regions \citep{williams07, mackey11}. In the case of M16, submillimeter dust polarization observations have shown that the magnetic field lines are aligned with the pillars, and perpendicular to and decoupled from the field of the surrounding photo-ionized cloud \citep{pattle18}.

The best-known example of such a region, M16, has recently been imaged using NIRSpec and MIRI, revealing an incredible level of detail in these ``Pillars of Creation'' \citep{hester96}.
\cmark
\cite{karim23} presented SOFIA and ground-based spectroscopic observations of this region. They infer that the ionized, atomic, and molecular phases are in pressure equilibrium if the atomic gas is magnetically supported. The dense clumps at the tops of the pillars are currently supported by the magnetic field, but are likely to collapse within the photoveaporation time scales.
\umark
The present study focuses on IC 1848, a similar region with a lower-UV-illumination, for which an extensive set of supporting FIR continuum and line observations from \emph{Spitzer}, SOFIA, and \emph{Herschel} already exists.

\begin{figure*}
   \sidecaption
   \includegraphics[trim=2cm 2.5cm 9cm 2.5cm, clip=true, width=12cm]{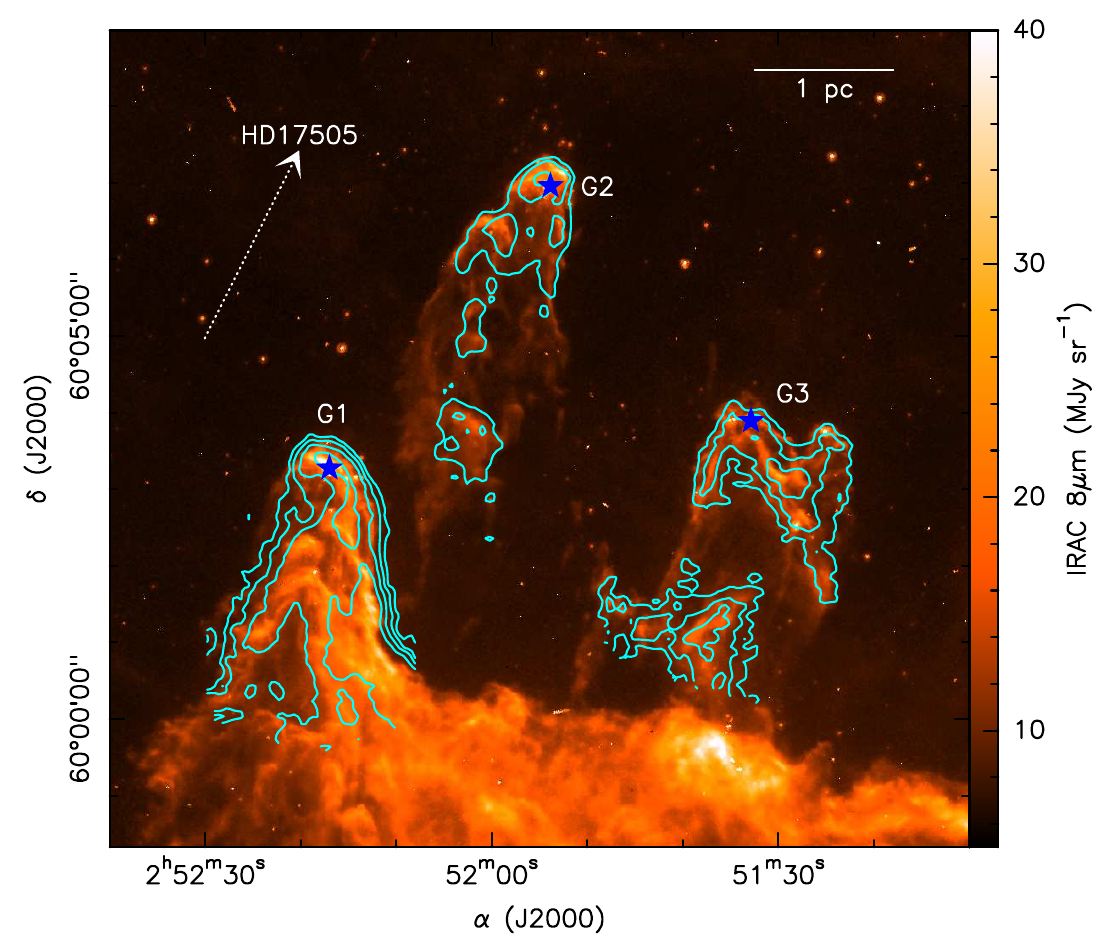}
   \caption{\emph{Spitzer}/IRAC 8 $\mu$m image of the three filaments in IC 1848 (G1–G3, left-to-right, respectively). Bright IRAC compact sources are clearly visible at the head of the filaments (white saturated spots in the image). Cyan contours show the [C\,\,{\sc ii}] 158~$\mu$m integrated line intensity observed with SOFIA/upGREAT. Contour levels are 8, 12, 18 and 27 K\,km\,s$^{-1}$, and typical 1$\sigma$ noise levels are 3.3, 1.8, and 2.4 K\,km\,s$^{-1}$ in G1--G3, respectively. Blue stars mark the location of the three brightest 2MASS/WISE sources at the heads of the filaments and the white arrow shows the direction toward the O6V ionizing star HD17505.} \label{irac}
 \end{figure*}

IC 1848 is a large HII region in Cassiopea, forming part of the W5 HII region-molecular cloud complex in the Perseus Arm, at a distance 1.9 kpc \citep{ishida70}, consistent within uncertainties with the recent distance estimate of 2075+44--42~pc to the Cas OB 6 association, formed by W3/W4/W5, derived from the \emph{Gaia} EDR3 release \citep{maiz22}. Past observations have revealed three bright rimmed clouds on the southern edge of the HII region \citep{thompson04} at a projected distance of ~13 pc from HD17505, the primary O6V ionizing star. These filaments, referred to as G1 – G3 (Figure~\ref{irac}), are associated with recent or ongoing star formation, as evidenced by the presence of submillimeter cores and multiple YSO’s identified in 2MASS images. This strongly suggests that UV illumination may have induced the collapse of the dense molecular cores at the head of the three clouds and \cite{thompson04} argue that the overall morphology of the IC~1848 clouds is reasonably consistent with radiative-driven implosion models of cometary globules.

A recent study of IC 1848 \citep{chauhan11} suggests that the filaments may have been caused by hydrodynamical instability of the ionization/shock front of the expanding H II region, as similar structures often show up in recent numerical simulations of the evolution of H II regions. These authors further hypothesize that this hydrodynamic instability mechanism makes a third mode of triggered star formation associated with H II regions, in addition to the two previously recognized mechanisms, collect-and-collapse of the shell accumulated around an expanding H II region and radiation-driven implosion of bright rimmed clouds that were pre-existing cloud clumps. Accurate characterization of the physical conditions in the gas, through observations combined with state-of-the-art Meudon PDR and Cloudy modeling, would provide good quantitative constraints for such models.

To characterize the physical conditions and the velocity field in the IC 1848 filaments, observations of [C\,{\sc ii}] 158 $\mu$m and [O\,{\sc i}] 63 and 145 $\mu$m fine structure lines were carried out using the upGREAT instrument on SOFIA. Full velocity-resolved [C\,{\sc ii}] maps of the three filaments were obtained, along with [O\,{\sc i}] observations at selected positions. 

\section{Observations}
\label{sec:obs}

During six flights between December 2019 and April 2022, all departing from Palmdale, CA, the three trunks G1 to G3 (Figure~\ref{irac}) were observed with the array spectrometer upGREAT\footnote{The German REceiver for Astronomy at Terahertz frequencies (upGREAT) is a development by the MPI f\"ur
Radioastronomie and the KOSMA/Universit\"at zu K\"oln, in cooperation with the DLR Institut f\"ur Optische Sensorsysteme.} \citep{Risacher2018} aboard SOFIA \citep{Young2012}.
All observations were carried out on the GREAT Consortium time (total flight time 7.5 hours, including all overheads). 
First, the filaments were mapped in fast total power on-the-fly mode with both polarizations of the 7-pixel low-frequency arrays (LFA) tuned to the carbon fine-structure line \CII\ $^3P_{3/2}-^3P_{1/2}$ 158 $\mu$m ($f_0 = 1900.53690$~GHz; \citealt{cooksy86}). The three mapped fields were over-sampled at 3\arcsec\ spacing, with 0.2~sec integration time per dump.
The high-frequency array (HFA), tuned to the \OI\ $^3P_1 - ^3P_2$ 63 $\mu$m ($f_0 = 4744.77749$~GHz; \citealt{zink91}) transition, was operated in parallel, but post-flight analysis revealed a too weak line intensity for quantitative analysis. Therefore, we decided to revisit the three strongest \CII\ condensations, in each filament, with targeted, deep(er) integrations in rather slow chopped mode (rate 0.626 Hz) for better stability. For these observations the LFA was operated in split mode, with the LFA-V tuned to
\OI\ $^3P_0 - ^3P_1$ 145\,\micron\ ($f_0 = 2060.06886$~GHz), while the LFA-H was still recording \CII. The HFA was running in parallel, but due to the frequency dependent array geometry, only the central pixels of the two arrays were co-aligned on the sky. 

Half-power main beam sizes ($\theta_{mb}$: 14\farcs1 for \CII, 13\arcsec\
for \OI\ 145~\micron\ and 6\farcs3 for \OI\ 63~\micron) and efficiencies (<$\eta_{mb}$> = 0.64 for LFAV, 0.66 for LFAH, and 0.68 for HFA, respectively), derived from
planet observations, were provided by the GREAT consortium. The instrument parameters are described in \cite{Risacher2018}. Both spectrometer arrays use NbN-based hot electron bolometer mixers, driven by either solid-state (LFA) or quantum cascade laser (HFA) local oscillator sources. The LFA operates 14 pixels, seven in each polarization, in a hexagonal arrangement with 31\farcs8 radial spacing around a central pixel. The HFA has the same
symmetry, using the V polarization only, with 13\farcs8 radial spacing. Co-alignment between the central pixels is generally better than 1\arcsec\  \citep{Risacher2018}. For this experiment we tracked on the central HFA pixel.
Pointing was established by the telescope operators on nearby optical reference stars, to an accuracy of 1-2". Prior to each flight series, the optical axis of the GREAT instrument had been aligned to these imagers by observations of planets.
Fast Fourier Transform Spectrometers (FFTS4G, updated from \citep{Klein2012}) offer 32k channels across the 0–4 GHz intermediate-frequency
bands.

Raw data were amplitude calibrated with the kosma {\it{kalibrate}} software package (version 2022.04), following \citep{Guan2012}. The data were corrected for atmospheric
extinction and calibrated in T$_{mb}$. The atmospheric transmission was smooth near the targets' velocities, except for the pointed observations of \OI\ 145\,\micron\ towards -G3 (in April 2022) that were made impossible because of strong blending with a prominent telluric ozone feature.

Further processing of the calibrated data was performed with the GILDAS software\footnote{http://www.iram.fr/IRAMFR/GILDAS}.
An initial inspection of the data cube was done in order to identify and eliminate unusable spectra with artefacts. Spectra with baseline ripples were identified by comparing the baseline noise with the radiometric noise deduced from the calibration and discarded.
Once in main beam temperature, the spectra were smoothed to a resolution between 0.1 and 0.15 km\,s$^{-1}$ and baselines of orders one or two were removed.

\begin{figure*}
   \centering
   \includegraphics[trim=2cm 2.5cm 7cm 2.5cm, clip=true, width=0.8\textwidth]{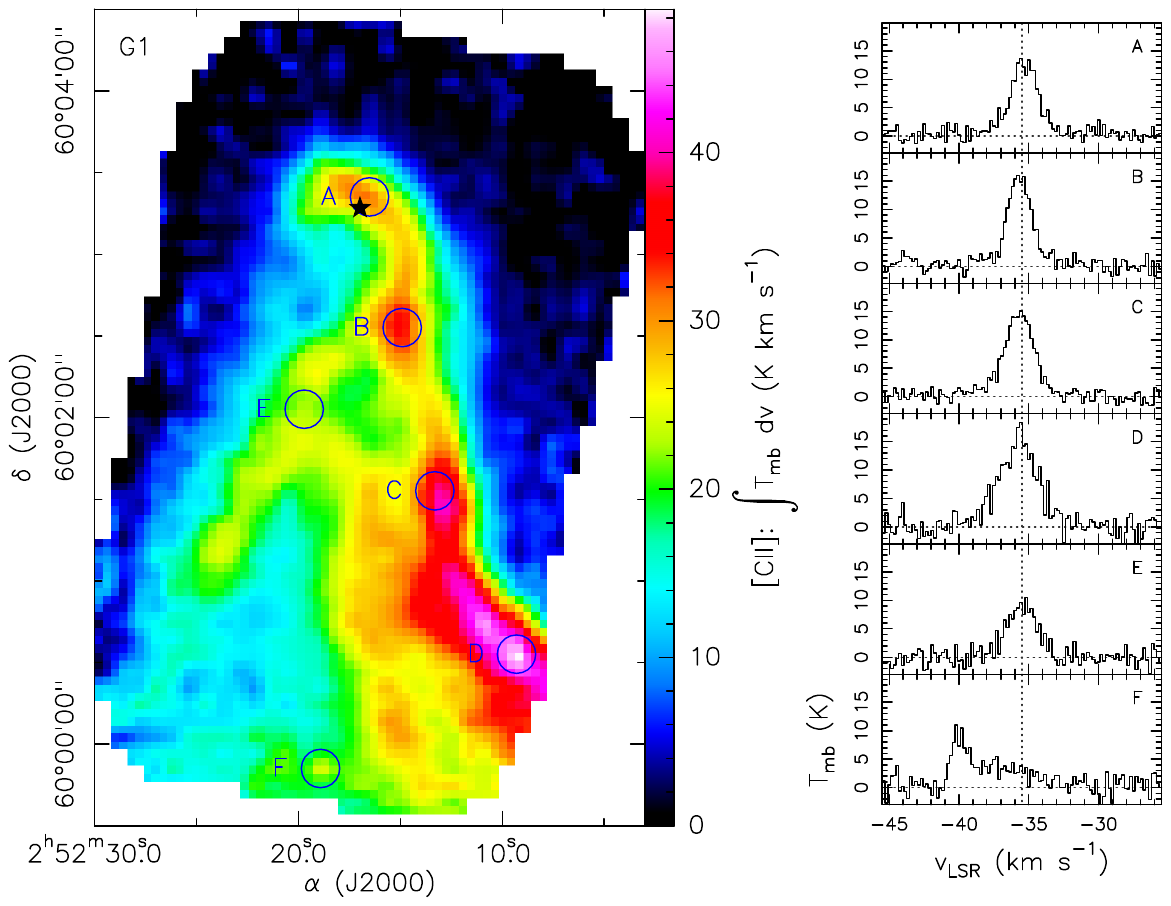} 

   \includegraphics[trim=2cm 2.0cm 5cm 1.5cm, clip=true, width=0.8\textwidth]{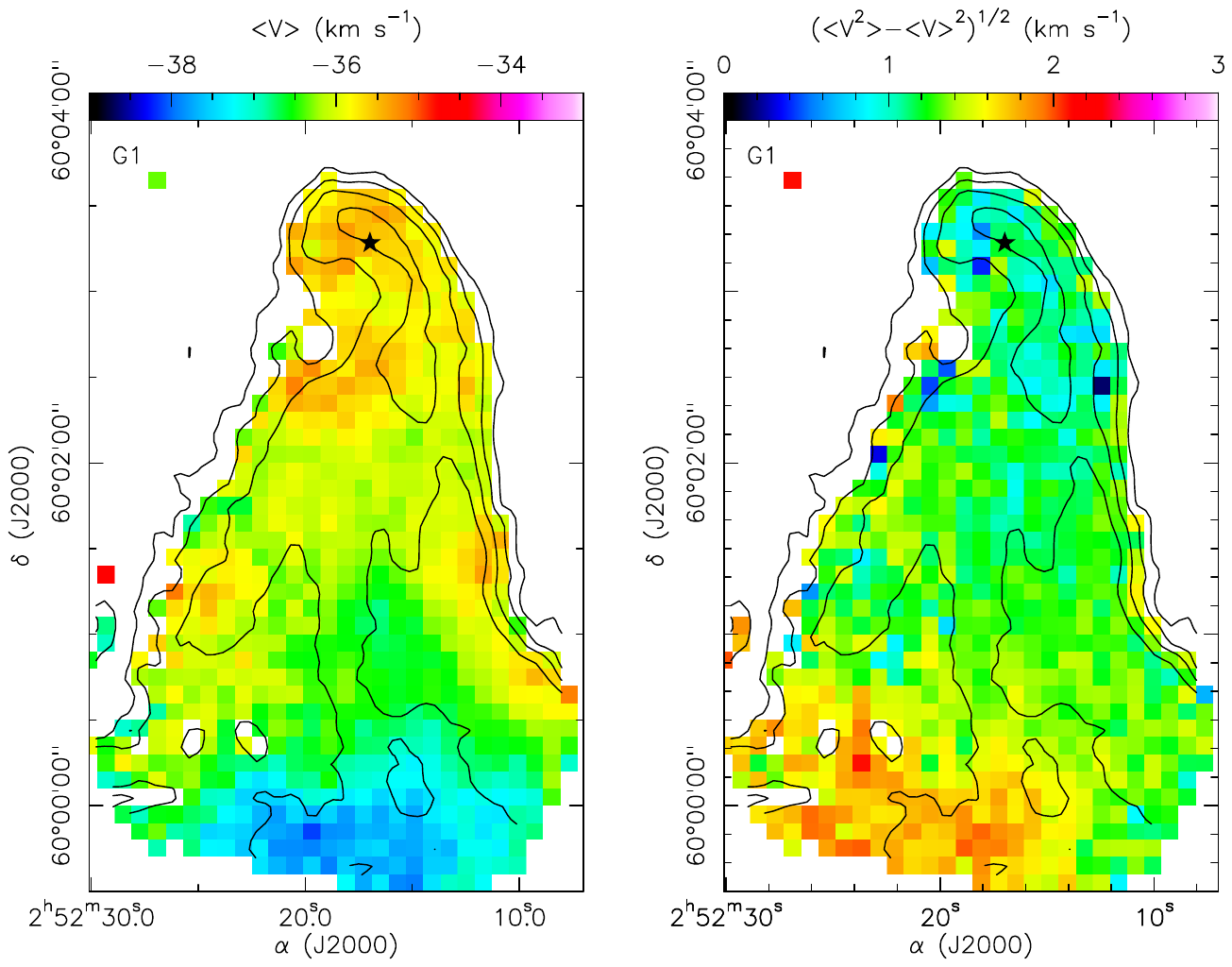} 
   \caption{(Top) Color image of the [C\,\,{\sc ii}] integrated line intensity in the --40 to --33.5 km\,s$^{-1}$ velocity range toward the G1 filament. The right panel shows spectra at selected positions labeled A--F in the image. Blue circles correspond to the FWHM SOFIA beam size at the [C\,\,{\sc ii}] frequency. (Bottom) Maps of the [C\,\,{\sc ii}] line center velocity and velocity dispersion toward the G1 filament (color images, left and right panels, respectively). Only pixels with [C\,\,{\sc ii}] intensities above 3.5$\sigma$ (11.6 K\,km\,s$^{-1}$) are shown. Black contours show the integrated line intensity, with the same contour levels as in Figure~1. Black stars mark the location of the bright IRAC compact source.} \label{momentsg1}
\end{figure*} 

\section{Results and discussion}
\label{sec:results}

\subsection{Morphology and kinematics}

Figure~\ref{irac} shows excellent correlation between the IRAC 8 $\mu$m continuum emission in IC~1848 (color image), likely dominated by the PAH emission, and the [C\,\,{\sc ii}] line emission (cyan contours). Bright IRAC point sources are located at the head of each pillar, which coincide with 2MASS point sources classified as YSO or YSO Candidate. Their 2MASS and WISE magnitudes are listed in Table~1. From their WISE colors \citep{koenig14, kang17} the sources at the head of the G1 and G2 filaments can be identified as Class I, while that at the head of the G3 filament is at the boundary between Class I and Class II.

To estimate bolometric luminosities of the embedded stars from their 2MASS magnitudes, we use eq. (5) of \cite{gonzalez09}. We use the $K_S$ magnitudes, which are least affected by extinction, and adopt the Vega K magnitude and absolute integrated flux from Table 3 of \cite{gonzalez09}). The resulting bolometric luminosities are 0.3--0.9 $L_{\odot}$ (Table 1). These values are not corrected for extinction.


\begin{table*}
\begin{center}  
\label{tab:wise}
\caption{WISE and 2MASS Magnitudes of the Point Sources at the Heads of the G1--G3 Filaments.-} 
\begin{tabular}{cccccccccccc}
\hline \hline 
  \rule[-3mm]{0mm}{8mm} $Pillar$ & $ 2MASS Source$ & $J$ & $H$ & $K_S$ & $w1$ & $w2$ & $w3$ & $w2-w3$ & $w1-w2$ & $Class$ & $L_{bol}$\\
  \hline 
  G1 & J02521697+6003170 & 16.498 & 14.94  & 14.546 & 10.895 & 8.659 & 4.961 & 3.698 & 2.236 & I    & 0.37 \\
  G2 & J02515380+6006581 & 15.934 & 13.625 & 11.776 & 8.883  & 7.018 & 4.889 & 2.129 & 1.865 & I    & 0.81 \\
  G3 & J02513283+6003542 & 13.005 & 11.523 & 10.447 & 9.261  & 8.112 & 5.562 & 2.550 & 1.149 & I/II & 0.86 \\
  \hline 
\end{tabular}
\end{center}
Note: Entries in the table are: 2MASS source identifier, 2MASS J, H, and K$_S$ magnitudes (1.235, 1.662, and 2.159 $\mu$m), WISE w1 -- w3 magnitudes (3.4, 4.6, and 12 $\mu$m), 
WISE colors, the resulting classification \citep{koenig14,kang17}, and an estimate of the bolometric luminosity (in solar luminosities) based on the $K_S$ magnitude \citep{gonzalez09}. We assume a distance of 1.9 kpc corresponding to the distance modulus of 11.39 mag, a Vega $K_S$-band magnitude of 0.043 and a flux of 1.1e-7 erg\,s$^{-1}$\,cm$^{-2}$ (Table 3 of \citealt{gonzalez09}).
\end{table*}

Figure~\ref{momentsg1}, top panel, shows the morphology of the [C\,\,{\sc ii}] line emission toward the brightest G1 filament. A bright, narrow ridge of emission is seen on the west side of the filament, with weaker emission extending toward the east. Example [C\,\,{\sc ii}]  spectra toward selected positions marked A--F are shown on the top-right. With the exception of the position F at the base of the filament, which shows a broad line with two velocity components, the remaining spectra show consistent velocities, with the head of the filament (position A) displaying slightly redshifted emission compared to the other positions, possibly owing to the radiation pressure exerted by the impinging UV photons or the rocket effect of the evaporating gas \citep{bertoldi90}, which causes the cloud to be accelerated away from the ionizing star. \cite{bertoldi90} estimates the characteristic rocket velocity as 10--5 km\,s$^{-1}$. This is higher than the $\simeq 1$~km\,s$^{-1}$ velocity difference between the surface and the center of the G1 filament (Figure~\ref{momentsg1}, lower-left), possibly due to projection effects. The [O\,{\sc i}] velocities at the head of the G1 filament are consistent with [C\,{\sc ii}] (Figure~A.6, positions G, H).

\begin{table*}
\begin{center}  
\label{tab:seds}
\caption{Molecular Hydrogen and [CII] Column Densities.} 
\begin{tabular}{ccccccccccc}
\hline \hline 
\rule[-3mm]{0mm}{8mm} Position & ($\Delta\alpha,\Delta\delta$) & $T_d$ & $\tau_{350}$ & $\beta$ & $N({\rm H_2})$ & $A_V$ & [C \sc{ii}] & $N$(C$^+)$ \\
\rule[-3mm]{0mm}{0mm} & (\arcsec) & (K) & & & (cm$^{-2}$) & & (K\,kms$^{-1}$) & (cm$^{-2}$) \\
  \hline 
  G1~A & ($-5.1,3.8$) & 34.7 & $1.1 \times 10^{-3}$ & 0.00 & --                   & --   & 24.3 & &                    \\ 
  G1~B & ($-17.1,-44.2$) & 28.5 & $8.3 \times 10^{-4}$ & 0.69 & $1.00 \times 10^{22}$ & 9.1  & 28.4 & $3.3 \times 10^{17}$ \\ 
  G1~C & ($-29.1,-104.2$) & 29.4 & $3.4 \times 10^{-4}$ & 0.89 & $4.16 \times 10^{21}$ & 3.8  & 32.2 & $3.8 \times 10^{17}$ \\ 
  G1~D & ($-59.1,-164.2$) & 27.3 & $3.6 \times 10^{-4}$ & 1.07 & $4.34 \times 10^{21}$ & 3.9  & 39.2 & $4.6 \times 10^{17}$ \\ 
  G1~E & ($18.9,-74.2$) & 27.9 & $1.0 \times 10^{-4}$ & 0.55 & $1.23 \times 10^{21}$ & 11.1 & 20.3 & $2.4 \times 10^{17}$ \\ 
  G1~F & ($12.9,-206.2$) & 28.0 & $3.6 \times 10^{-4}$ & 0.85 & $4.34 \times 10^{21}$ & 3.9  & 18.5 & $2.2 \times 10^{17}$ \\ 
\hline
  G2~A & (3.5,7.3) & 32.5 & $9.6 \times 10^{-4}$ & 0.36 & $1.16 \times 10^{22}$ & 10.5 & 15.5 & $1.8 \times 10^{17}$ \\ 
  G2~B & ($39.5,-34.7$) & 30.1 & $6.1 \times 10^{-4}$ & 0.23 & $7.35 \times 10^{21}$ & 6.6  & 13.1 & $1.5 \times 10^{17}$ \\ 
  G2~C & ($-8.5,-34.7$) & 26.0 & $4.1 \times 10^{-4}$ & 0.90 & $4.90 \times 10^{21}$ & 4.4  & 10.2 & $1.2 \times 10^{17}$ \\ 
  G2~D & ($63.5,-112.7$) & 24.7 & $2.7 \times 10^{-4}$ & 0.72 & $3.25 \times 10^{21}$ & 2.9  & 8.3  & $1.0 \times 10^{17}$ \\ 
  G2~E & ($63.5,-190.7$) & 26.3 & $2.4 \times 10^{-4}$ & 0.67 & $2.89 \times 10^{21}$ & 2.6  & 11.0 & $1.3 \times 10^{17}$ \\ 
\hline
  G3~A & ($-2.9,-13.6$) & 28.8 & $7.0 \times 10^{-4}$ & 0.51 & $8.43 \times 10^{21}$ & 7.6  & 15.3 & $1.8 \times 10^{17}$ \\ 
  G3~B & ($-32.9,-43.6$) & 27.6 & $4.7 \times 10^{-4}$ & 0.80 & $5.70 \times 10^{21}$ & 5.2  & 19.2 & $2.2 \times 10^{17}$ \\ 
  G3~C & ($33.1,-49.6$) & 27.9 & $2.9 \times 10^{-4}$ & 0.76 & $3.50 \times 10^{21}$ & 3.2  & 10.3 & $1.2 \times 10^{17}$ \\ 
  G3~D & ($-56.9,-115.6$) & 29.7 & $2.1 \times 10^{-4}$ & 0.57 & $2.55 \times 10^{21}$ & 2.3  & 8.6  & $1.0 \times 10^{17}$ \\ 
  G3~E & ($39.1,-169.6$) & 29.2 & $2.3 \times 10^{-4}$ & 0.80 & $2.81 \times 10^{21}$ & 2.5  & 18.0 & $2.1 \times 10^{17}$ \\ 
\hline
\end{tabular}
\end{center}

Note: Entries in the table are: position (see Figure~2, top panel), offsets from the image center, best fit dust temperature, 350~$\mu$m optical depth, grain emissivity exponent, H$_2$ column density in the 36\farcs7 SPIRE beam at 500~$\mu$m and the corresponding visual extinction, [C\,{\sc ii}] integrated line intensity in a 36\farcs7 beam, and the corresponding C$^+$ column density. Image center positions (J2000) are: (02:52:17.19, +60:03:17.22) for G1, (02:51:53.91, +60:06:56.97) for G2, and (02:51:32.90, +60:03:53.34) for G3.
\end{table*}

To investigate further the kinematics of the gas across the filament, we have computed the first and second moment maps of the [C\,\,{\sc ii}] line emission. The bottom-left panel shows the mean [C\,\,{\sc ii}] velocity computed as $\overline{v} = \sum v_i T_i / \sum T_i$, and the bottom-right panel shows the velocity dispersion, $\sigma_v = \sqrt{\overline{v^2}-\overline{v}^2}$. The mean line velocity increases by up to $\simeq 1$~km\,s$^{-1}$ at the head and near the edges compared to the center of the filament, while the velocity dispersion remains uniform within the observational uncertainties. A hint of this velocity shift can also be seen in the position A spectrum in Figure~\ref{momentsg1}. The base of the G1 filament is characterized by a significantly lower mean velocity and higher velocity dispersion, owing to the presence of a distinct velocity component at $-40$~km\,s$^{-1}$ (Figure~\ref{momentsg1}, position F), which is likely a part of the extended molecular shell that can be seen in the IRAC image (Figure~\ref{irac}).
\cmark
Outside of this region, the velocity dispersion is quite uniform along the filament, in contrast to Pillar 2 in M16, where the dispersion increases toward the tail (see Fig.~8 of \citealt{karim23}).  This is likely due to the difference in the UV illumination between the two regios. The three M16 pillars are illuminated by a handful of bright O5--7 stars located at a $\sim 3$~pc projected distance \citep{hillebrand93}. \cite{karim23} estimated the FUV field strength toward the M16 pillars is in the range 1500 -- 2500 Habing units. Consequently, the UV photons have smaller influence on the kinematics of the gas and the resulting degree of turbulence in IC~1848 compared to M16.
\umark

Moment maps of the G2 and G3 filaments are shown in Figures~\ref{momentsg2}--\ref{momentsg3} in the Appendix. Filament G2 shows an elliptical morphology, with the west side (Figure~\ref{momentsg2}, positions A and C) significantly redshifted compared to the east side (positions B, D, and E). There is some evidence for lower velocity dispersions on the west side, but the uncertainties are large. Filament G3 splits into two distinct parts, the head and the base, which display large differences in the mean velocity and velocity dispersion (Figure \ref{momentsg3} in the Appendix). The [C\,\,{\sc ii}] spectrum at the base is double peaked, indicating the presence of a separate velocity component at about $-43$~km\,s$^{-1}$.

The kinematic structure of the three filaments is further displayed in the velocity channel maps (Figures~\ref{chang1}--\ref{chang3} in the Appendix, filaments G1--G3, respectively). Based on the morphology of the H$\alpha$ emission, \cite{thompson04} argued that although the three clouds lie close together on the sky, this arrangement may be a chance superposition, with G1 located behind the illuminating star and G2/G3 in front. Such a geometry could explain the $\sim 5$ km\,s$^{-1}$ velocity differences between G1 and the other two clouds.
\cmark
However, the observed velocity gradients observed along all three filaments, with the [C\,\,{\sc ii}] emission at the head being redshifted with respect to the tail, would seem to indicate that the filaments are collocated on the front side of the illuminating star (see, e.g., \citealt{bertoldi90, pound98}).
As discussed in Sect.~3.4, the observed [C\,{\sc ii}]/[O\,{\sc i}] 63~$\mu$m line intensity ratio does not seem consistent with back-side illuminated PDR models of G2 and G3. The exact geometry of the three filaments with respect to the illuminating star remains, therefore, ill-defined.
\umark

\begin{figure*}
   \sidecaption
   \includegraphics[trim=2.0cm 3.5cm 6cm 2.5cm, clip=true, width=12cm]{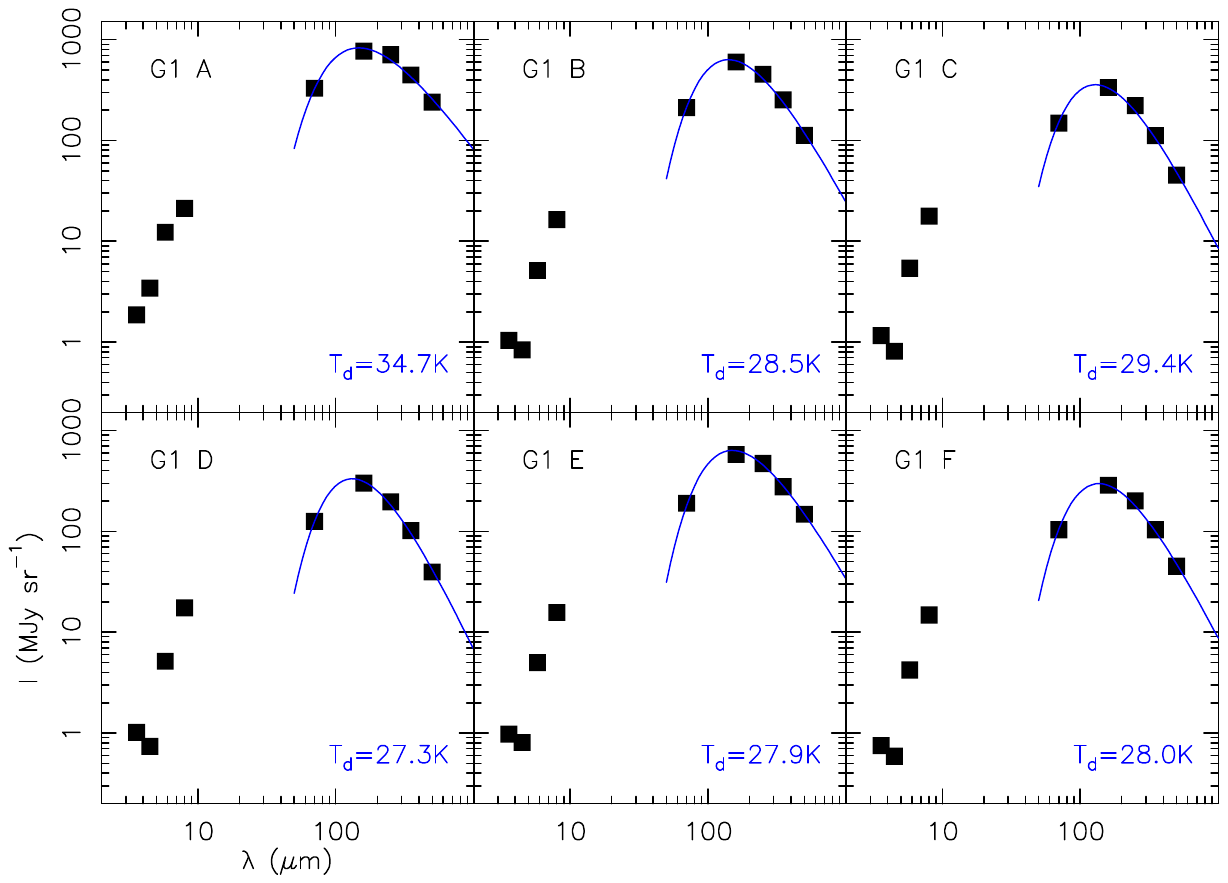} 
   \caption{Spectral energy distributions toward positions A--F in the G1 filament based on \emph{Spitzer} and \emph{Herschel} observations. Blue curves show the modified blackbody fits to the \emph{Herschel} PACS and SPIRE fluxes from which the dust temperatures and 350~$\mu$m optical depths used for the column density determination are derived. Typical flux calibration uncertainties are $\sim 5$\%.} \label{sedg1}
\end{figure*}

\subsection{Hydrogen column densities}

In addition to the SOFIA and \emph{Spitzer} data, far-infrared continuum images of IC 1848, obtained using PACS and SPIRE instruments, are available in the \emph{Herschel} Science Archive, which allow characterization of the continuum spectral energy distribution (SED) over a broad wavelength range to constrain the dust temperature and the H$_2$ column density required for the PDR modeling (Sec.~3.4). Figure~\ref{sedg1} shows the continuum SEDs at positions A--F in the G1 filament. All continuum images have been convolved to the 36\farcs7 resolution (full width at half maximum, FWHM) of the longest-wavelength 500~$\mu$m SPIRE channel. Blue curves show single-component modified blackbody fits, $I_\nu = B_\nu (T_d) \times [1 - \exp(-\tau_{350} \times (350 \mu {\rm m} / \lambda)^\beta)]$, to the PACS and SPIRE data only. Best fit parameters, the dust temperature ($T_d$), the 350~$\mu$m optical depth ($\tau_{350}$), and the grain emissivity exponent ($\beta$) are listed in Table~2.

The SED toward position A is well described by a scaled blackbody ($\beta = 0$). This is consistent with the flux, even at the longest SPIRE wavelengths, being dominated by the stellar source seen in the IRAC images rather than dust emission. The remaining positions have steeper dust emission spectra characterized by average dust temperatures of $\sim 28$~K, low 350~$\mu$m optical depths, and shallow values of $\beta \sim 0.8$.

The H$_2$ column densities in a 36\farcs7 beam are computed using the formula $N({\rm H_2}) = 2 a \rho R_{gd} / 3 m_{\rm H} \times \tau_{350} / Q_{350}$ \citep{lis90}, where $a = 0.1\, \mu$m is the mean grain radius, $\rho = 3$~g\,cm$^{-3}$ is the mean density, $R_{gd} = 100$ is the gas-to-dust mass ratio, $m_{\rm H}$ is the hydrogen mass, and $Q_{350}$ is the 350~$\mu$m grain emissivity coefficient. Extrapolating the 125 $\mu$m grain emissivity coefficient of \cite{hilde83} ($7.5 \times 10^{-4}$) with a $\nu^2$ frequency dependence gives $Q_{350} = 1 \times 10^{-4}$, corresponding to the grain mass opacity coefficient $\kappa_{350} = 3 Q / 4 / a /\rho / R_{dg} = 0.025$~cm$^2$\,g$^{-1}$, including the gas-to-dust ratio (see \citealt{lis98}). The resulting H$_2$ column  densities are listed in Table~2 together with the corresponding visual extinction values, computed assuming the conversion factor of $N_{\rm H} = 2.21 \times 10^{21}$~cm$^{-21} ~ A_V ({\rm mag})$ \citep{guver09}.
The typical visual extinction values characterizing the [C\,{\sc ii}] emitting gas are moderate, ranging from $A_V \sim 10$ near the head of the filament to $\sim 4$ at the base. The characteristic width of the G1 filament is $\sim 90^{\prime\prime}$, corresponding to 0.8~pc, or $2.6 \times 10^{18}$~cm at a distance of 1.9~kpc. Assuming the same depth along the line of sight, the H$_2$ column densities toward the head of the filament correspond to a mean volume density of $\sim 4500$~cm$^{-3}$. This should be considered a lower limit, as narrower filamentary structures can be seen in the \CII\ and dust continuum images. These values are used as input for the PDR models described below.

Modified blackbody fits toward positions A--E in the G2 and G3 filaments are shown in Figure~\ref{sedg2g3} in the Appendix and best fit parameters are listed in Table~2. The dust temperatures, H$_2$ column densities, and the corresponding visual extinction values are comparable to those in the G1 filament. The dust temperature, $A_V$, and average density estimates derived here are in a reasonable agreement with those derived by \cite{thompson04} based on SCUBA and IRAS HIRES fluxes (see Table 5 of \citealt{thompson04}).

\begin{figure*}
   \sidecaption
   \includegraphics[trim=2.5cm 3.0cm 6.5cm 7.5cm, clip=true, width=12cm]{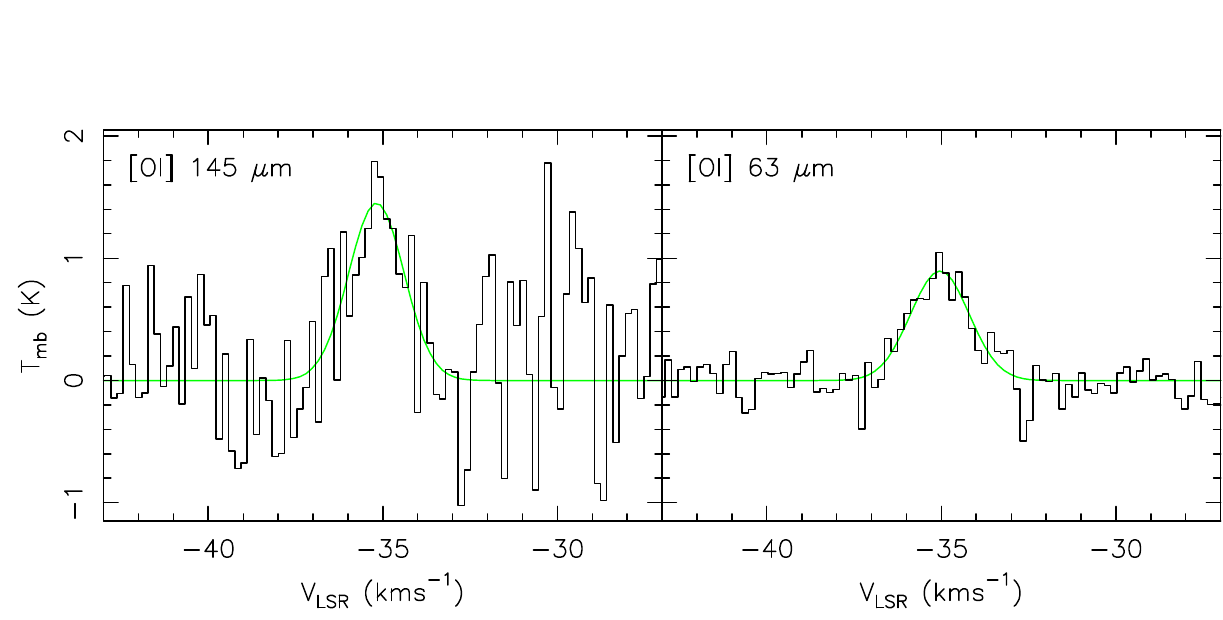} 
   \caption{Equally weighted average \OI\ spectra toward the head of the G1 filament (within the white rectangle in Figure~\ref{oxygen}). Averages include \OI\ 63~$\mu$m spectra at 14 individual positions and \OI\ 145~$\mu$m spectra at 3 individual positions. The excess noise on the right hand side of the \OI\ 145~\micron\ spectrum is due to increased atmospheric opacity of a nearby telluric ozone line.} \label{oig1head}
\end{figure*}

\begin{table*}
\begin{center}  
\label{tab:oxy}
\caption{[C\,{\sc ii}] and [O\,{\sc i}] Line Intensities at Selected Positions.} 
\begin{tabular}{cccccc}
\hline \hline 
\rule[-3mm]{0mm}{8mm} Position & ($\Delta \alpha, \Delta \delta$) & [C\,{\sc ii}] &  [O\,{\sc i}] 63 $\mu$m &  [O\,{\sc i}] 145 $\mu$m & [O\,{\sc i}]/[C\,{\sc ii}] \\
\rule[-3mm]{0mm}{0mm} & (\arcsec) & (K\,kms$^{-1}$) & (K\,kms$^{-1}$) & (K\,kms$^{-1}$) \\
\hline 
  G1~G & ($1.0,5.0$) & $29.5\pm 0.91$ & $4.02\pm0.43$ &  --  & 0.14 \\     
  G1~H & ($-12.4,5.4$) & $26.2\pm0.85$ & $2.88\pm0.40$ &  --  & 0.11 \\     
  G1~I & ($-20.5,-33.6$) & $24.1\pm0.84$ & $2.51\pm0.44$ &  --  & 0.10 \\     
  G1~J & ($-13.0,-45.0$) & $31.8\pm0.77$ & $2.91\pm0.41$ &  --  & 0.09 \\     
  G1~K & ($-26.4,-44.7$) & $20.5\pm0.85$ & $2.09\pm0.39$ &  --  & 0.10 \\     
  G1~L & ($-23.1,-56.2$) & $27.0\pm0.81$ & $2.04\pm0.44$ &  --  & 0.08 \\     
  \hline 
  Average & & $26.5\pm1.5$ &  $2.74\pm0.30$   &  & 0.10  \\
  \hline 
  G1 Head & & $19.9\pm0.9$  & $1.97\pm0.13$ & --            & 0.10 \\
  \hline 
  G1~M & ($4.6,-12.3$) & $13.3\pm0.85$ &  --  & $4.41\pm0.98$ & 0.33 \\     
  G1~N & ($-14.0,-21.0$) & $21.7\pm0.80$ &  --  & $3.62\pm0.69$ & 0.17 \\     
  G1~O & ($-30.4,-117.3$) & $34.2\pm0.95$ &  --  & $3.81\pm0.93$ & 0.11 \\     
  \hline 
  Average & & $23.1\pm6$ &               & $3.95\pm0.24$  & 0.20 \\
  \hline
  G1 Head & & $19.9\pm0.9$  &               & $2.93\pm0.36$ & 0.15 \\
  \hline 
  G2~F & (10.9,4.9) & $14.0\pm1.04$ & $1.53\pm0.23$ &  --  & 0.11 \\     
  G2~G & ($-2.0,4.3$) & $19.4\pm0.83$ & $1.49\pm0.23$ &  --  & 0.08 \\     
  G2~H & ($4.1,16.4$) & $17.7\pm0.87$ & $1.72\pm0.22$ &  --  & 0.10 \\     
  G2~I & ($33.1,-25.6$) & $12.9\pm0.87$ & $2.02\pm0.30$ &  --  & 0.16 \\     
  \hline 
  Average & & $15.6\pm1.3$    &     $1.59\pm0.14$          & & 0.10 \\
  \hline 
\end{tabular}
\end{center}

Note: Entries in the table are: position (see Figure~\ref{oxygen}, left panel), offsets from the image center, the [O\,{\sc i}\ and [C\,{\sc ii}] integrated line intensities, and the corresponding [O\,{\sc i}] to [C\,{\sc ii}] intensity ratio. \OI\ line intensities are computed in the velocity range --37.2 to --32.9 km\,s$^{-1}$ for G1 and -44.1 to --37.9 km\,s$^{-1}$ for G2, corresponding to two FWHM line widths of the corresponding average \OI\ 63~$\mu$m spectra.
\cmark
The uncertainties for the line intensities at individual positions and the average \OI\ 145~$\mu$m line toward the G1 Head are computed from the spectra, while those listed for the remaining averages are uncertainties of the mean ($\sigma/\sqrt(N)$, where $\sigma$ is the standard deviation of the ensamble and $N$ is the number of measurements).
\umark
Image center positions (J2000) are: (02:52:17.19,+60:03:17.22) for G1, and (02:51:53.91, +60:06:56.97) for G2.
\end{table*}
In addition to the H$_2$ column densities derived from dust SED fits, Table~2 lists the [C\,{\sc ii}] line intensities and the corresponding C$^+$ column densities at the same positions derived using equations (26) and (43) of \cite{goldsmith12}. PDR models (Figure~\ref{fig:pdr}) suggest that the [C\,{\sc ii}] emission originates in the outer $A_V \sim 1.5$ molecular layer characterized by gas temperatures between $\sim 50$ and 160~K. Consequently, we use an average temperature of 100~K and an average H$_2$ density of 4500~cm$^{-3}$ derived above in the C$^+$ column density calculations. Typical C$^+$ column densities in the G2 and G3 filaments are of order $1 - 2 \times 10^{17}$~cm$^{-2}$, while those in the G1 filament are a factor of $\sim 2$ higher (Table~2).  Assuming a C$^+$ abundance of $\sim 2.5 \times 10^{-4}$ with respect to H$_2$ in the outer molecular layer, as suggested by PDR models, the [C\,{\sc ii}] emission traces on average $\sim 17$\% of the total H$_2$ column density derived from the dust SED fits.
\cmark
The C$^+$ column densities in IC~1848 are toward the low end of the values derived by \cite{karim23} in the M16 pillars, $\sim 2\times10^{17} - 4.7 \times 10^{18}$ cm$^{-2}$. This is once again likely due to the difference in the FUV illumination between the two regions.
\umark

\begin{figure}
   \centering

   \includegraphics[trim=0cm 0cm 0cm 0cm, clip=true, width=1.0\columnwidth]{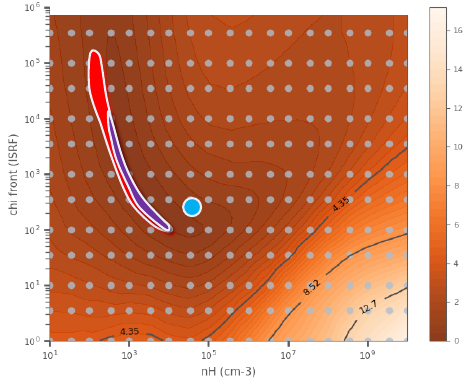} 
   \caption{$\chi^2$ values for a grid of isochoric ISMDB models compared to the observed fine structure line intensities in the G1 filament. Blue circle and the elongated red area show best fits, respectively, to the \CII/\OI\ 145~$\mu$m data at positions M--O and \CII/\OI\ 63~$\mu$m data at positions G--L.  The elongated magenta area shows the best fit to all data with \OI\ 145~$\mu$m line intensities scaled down by a factor of 3, as described in the text. All models assume normal incidence. Increasing the incidence angle to 60$^\circ$ provides comparable fit, but shifts the best-fit solution to higher densities and lower $\chi$ values (magenta area).} \label{ismdb}
\end{figure} 

\subsection{Atomic oxygen observations}

In addition to the fully-sampled [C\,{\sc ii}] maps, the 63~$\mu$m and 145~$\mu$m [O\,{\sc i}] fine structure lines were observed at selected positions. [O\,{\sc i}] spectra toward the G1 filament with detections above 4$\sigma$ level for the integrated line intensity are shown in Figure~\ref{oxygen} (upper panel) in the Appendix. The 63~$\mu$m line is detected at 6 positions, labeled G--L, while the 145~$\mu$m line is detected at 3 positions, labeled M--O (see Figure~\ref{oxygen}). The \OI\ integrated line intensities are listed in Table~3, together with [C\,{\sc ii}] line intensities at the same positions extracted from the fully-sampled map.

The average [O\,{\sc i}]~63~$\mu$m/[C\,{\sc ii}] intensity ratio (in K\,km\,s$^{-1}$) for positions G--L is $\sim 0.10$, while the average [O\,{\sc i}]~145~$\mu$m/[C\,{\sc ii}] intensity ratio for positions M--O is $\sim 0.20$. This suggests that the [O\,{\sc i}]~145~$\mu$m line is typically stronger than the 63~$\mu$m line.The [O\,{\sc i}]~63~$\mu$m line is detected at four positions in the G2 filament (see Table~1 and Figure.~\ref{oxygen}, lower panel), and is not detected in the G3 filament. The [O\,{\sc i}]~63~$\mu$m/[C\,{\sc ii}] intensity ratios in the G2 filament are in the same range as those observed in G1.

Given the sparse and limited spatial coverage, limited signal-to-noise ratio of the \OI\ 145~$\mu$m spectra, and the difference in the beam size at the frequencies of the two oxygen lines, the [O\,{\sc i}]~63~$\mu$m/145~$\mu$m intensity ratio cannot be determined with high accuracy. However, positions N and I are located nearby, 14\arcsec\ apart along the ridge of the [C\,{\sc ii}] emission and the [O\,{\sc i}]~63~$\mu$m/145~$\mu$m intensity ratio at these positions in $\sim 0.7$.
An independent estimate of the [O\,{\sc i}]~63~$\mu$m/145~$\mu$m is obtained by averaging \OI\ spectra near the head of the G1 filament, within the white rectangle in Figure~\ref{oxygen}. The \OI\ 63 and 145\,$\mu$m lines are both detected with a signal-to-noise ratio of 14 and 8, respectively. The resulting \OI\ line intensities are reported in Table~3 (lines labeled ``G1 Head'') together with the average \CII\ integrated line intensity within the white rectangle. The [O\,{\sc i}]~63~$\mu$m/145~$\mu$m ratio is 0.67, consistent with the values derived above, further supporting the conclusion that the \OI\ 63~$\mu$m line is weaker than the 145~$\mu$m line.

\subsection{PDR modeling}

To model the fine structure line emission in IC~1848 we use the Meudon PDR code, a publicly available, state-of-the-art astrochemistry code \citep{lepetit06}. Our objective is to verify whether the observed line intensities are consistent with the geometry and the UV illumination of IC1848.  The Meudon code is well suited for our analysis, since it is one of very few astrochemistry codes that computes self-consistently the radiative transfer, the chemistry both in the gas phase and on grain surfaces, as well as the thermal processes and thermal balance at each point within the cloud.  The code simulates a stationary plane-parallel slab of gas and dust illuminated on both sides by the radiation field from the far UV (FUV) to the radio domain. This radiation field is discretized on a wavelength grid (typically 500,000 points) and the radiative transfer equation is solved at each wavelength, and at each position in the cloud, considering (a) absorption in the continuum by the distribution of grains and by the ionization of atoms such as carbon, (b) absorption in the atomic lines of hydrogen, the Lyman and Werner transitions of H$_2$ as well as the pre-dissociating lines of CO isotopologues (in total more than 30,000 UV absorption lines are considered), (c) backscattering by dust, and (d) re-emission of photons as the fluorescent UV photons emitted by H and H$_2$ (see \citealt{goico07}).

\begin{table*}
\begin{center}  
\label{tab:pdrmodel}
\caption{PDR Model Results.} 
\begin{tabular}{ccccccccccc}
\hline \hline 
  \rule[-3mm]{0mm}{8mm} Model & $A_V$ & Star &  $D$ & $\theta$ & $n_{\rm H}$ & $I$([C\,{\sc ii}]) & $I$([O\,{\sc i}] 63) & $I$([O\,{\sc i}] 145) & 
  $I$([O\,{\sc i}] 63)/ & $I$([O\,{\sc i}] 145)/ \\
  \rule[-3mm]{0mm}{0mm} & & & (pc) & (deg) & (cm$^{-3}$)& (K\,km\,s$^{-1}$) & (K\,km\,s$^{-1}$) & (K\,km\,s$^{-1}$) &
  $I$([C\,{\sc ii}]) &  $I$([C\,{\sc ii}]) \\
\hline 
  1027 & 10 & O 5 V & $-13$ &  0 & $9 \times 10^3$    &  26.43 & 1.74 & 0.53 & 0.066 & 0.020 \\
  1028 & 10 & O 5 V & $13$  &  0 & $9 \times 10^3$    &  20.51 & 0.50 & 0.53 & 0.025 & 0.026 \\
  1029 & 10 & O 5 V & $-15$ & 30 & $9 \times 10^3$    &  26.04 & 1.33 & 0.40 & 0.051 & 0.015 \\
  1030 & 10 & O 5 V & $15$  & 30 & $9 \times 10^3$    &  19.84 & 0.37 & 0.40 & 0.018 & 0.020 \\
  1031 & 10 & O 5 V & $-26$ & 60 & $9 \times 10^3$    &  21.88 & 0.41 & 0.11 & 0.019 & 0.005 \\
  1032 & 10 & O 5 V & $26$  & 60 & $9 \times 10^3$    &  15.81 & 0.10 & 0.11 & 0.006 & 0.007 \\
  1033 & 10 & O 5 V & $-15$ & 30 & $4.5 \times 10^4$  &  22.48 & 2.35 & 0.69 & 0.105 & 0.031 \\
  1034 & 10 & O 5 V & $15$  & 30 & $4.5 \times 10^4$  &  20.85 & 1.42 & 0.69 & 0.068 & 0.033 \\
  \hline
  Observed & -- & -- & -- & -- &  -- & 13--32 & 1.4--4.0 & 3.6--4.4 & 0.08--0.14 & 0.11--0.33 \\
  \hline
\end{tabular}
\end{center}
Note: Entries in the table are: model number, stellar type, distance (pc; negative and positive values imply front-side and back-side illumination, respectively), the [O\,{\sc i}] 63~$\mu$m, [O\,{\sc i}] 145~$\mu$m, and [C\,{\sc ii}] integrated line intensities (K\,km\,s$^{-1}$), the [O\,{\sc i}] 63 [C\,{\sc ii}]  line ratio, and the [O\,{\sc i}]/145~$\mu$m to [C\,{\sc ii}] line ratio. All models assume a turbulent velocity dispersion of 0.9~km\,s$^{-1}$, as suggested by the observed [C\,{\sc ii}] line widths and face-on illumination.

\end{table*}

This detailed solution of the radiative transfer equation gives the spectral energy density at all wavelengths and positions within the cloud, which is then used to compute locally precise photo-destruction rates by direct integration of the photo-absorption cross-sections \citep{heays17} with the monochromatic energy density. This means that photo-destruction rates include the UV shielding of molecules by H Lyman transitions and H$_2$ Lyman and Werner bands. So, whereas most of other astrochemical codes rely on basic approximations (simple exponential fits) to treat absorption of UV photons, here, thanks to the detailed treatment of the photo-destruction processes, the destruction rates of molecules are determined directly, and so their densities as a function of the position are computed more precisely.

For the chemistry, the code considers approximately 200 chemical species linked by a chemical network of $\sim 2000$ chemical reactions. In the gas phase, in addition to the standard exothermic reactions that constitute the bulk of chemical pathways, state-to-state reactions are considered for key endothermic reactions.  For the grain surface chemistry, a power law grain size distribution is considered \citep{mathis77}. The temperature for grains in each size bin is computed balancing absorption and emission of photons. 
The processes that are considered are the adsorption from the gas, reactions in the bulk and on the surface considering thermal and tunneling effects, desorption processes (thermal and non-thermal: photo-desorption, cosmic ray desorption, chemical desorption, i.e., release of molecules into the gas phase due to the exothermicity of some reactions). The Meudon PDR code has another strong advantage over most other astrochemistry codes: it computes very precisely the formation of H$_2$ on dust grains. Indeed, it can either consider Langmuir-Hinshelwood and Eley-Rideal processes with the rate equation formalism \citep{lebourlot12}, or even consider stochastic processes on dust \citep{lepetit09, bron14, bron16}.

\subsubsection{ISMDB models}

A particularly useful online tool is the InterStellar Medium DataBase\footnote{ISMDB; http://ismdb.obspm.fr}, which allows quick and easy access to pre-calculated theoretical models for various ISM conditions and objects, including models produced by the Meudon PDR code. Presently, ISMDB gives access to several thousand PDR models covering a wide range of astrophysical conditions. To prepare and interpret observations, ISMDB provides an inverse search service to allow queries on observables to find the best models that match observations and gives access to line intensity maps in a parameter space (e.g., the total hydrogen nuclei volume density $n_{\rm H} = n({\rm H}) + 2 \times n({\rm H_2})$, the ISRF scaling factor $\chi$ in Mathis units\footnote{See Sect. 3.2.1 of the Meudon PDR Code Documentation; https://data.obspm.fr/ism/files/PDRDocumentation/PDRDoc154.pdf}, \citealt{mathis83}). Once the best fit model is found, model results can be downloaded without the need for running the code on a local machine.

To analyze the observed fine structure line intensities in IC1848, we use the ISMDB inverse search tool based on a grid of isochoric (constant density) models computed using the latest public version of the Meudon PDR code (1.5.4). We assume front side illumination (enhanced illumination of side of cloud facing the observer), and the total thickness of the cloud corresponding to $A_V = 10$~mag, as derived above from the SED analysis. We note that the exact value of $A_V$ is not important for the computations, as the fine structure line emission originates from the surface of the PDR (see Figure~\ref{fig:pdr} below). We have verified that models with $A_V = 5$ and 10 mag predict essentially the same line intensities. The two key parameters varied in the fit are the gas density and the radiation field scaling intensity factor, $\chi$, with the observed fine structure line intensities being input observables.
We compute averages of the individual \CII\ and \OI\ 63~$\mu$m line intensities at positions G--L, and of the \CII\ and \OI\ 145~$\mu$m line intensities at positions M--O and use these values as input for the ISMDB fit, assuming a 30\%\ uncertainty as an estimate of both statistical and calibration uncertainties.

The $\chi^2$ for the ISMDB fit to the combined data set (Figure~\ref{ismdb}; color map and thin black contours) shows a broad, shallow distribution with no unique solution. Since we do not have observations of the two \OI\ fine structure lines at the same positions, we then fit separately the combinations of \CII/\OI\ 63~$\mu$m at positions G--L and \CII/\OI\ 145~$\mu$m at positions M--O. A fit to the observed \CII\ and  \OI\ 145~$\mu$m line intensities at positions M--O gives a unique solution with $n_{\rm H} \simeq 3 \times 10^4$~cm$^{-3}$ and $\chi \simeq 200 -250$ (blue dot), while the observed \CII\ and  \OI\ 63~$\mu$m line intensities at positions G--L are consistent with a range of parameters within the elongated red area. The two regions do not overlap; there is no unique solution that is consistent with both data sets. The \OI\ 145~$\mu$m observations suggest higher gas densities compared to the \OI\ 63~$\mu$m observations. For $\chi = 200 - 250$, the best fit for positions G--L has a density $n_{\rm H} \simeq 1 - 3 \times 10^3$~cm$^{-3}$, an order of magnitude lower than the value derived from the \OI\ 145~$\mu$m fit. These results show that the observed \OI\ 63~$\mu$m/145~$\mu$m ratio cannot be explained easily with the PDR models. To bring the two solutions into agreement, we must lower the \OI\ 145~$\mu$m line intensity by about a factor of 3. The best fit to the combined data set with such scaling is shown as the narrow magenta elongated area.
We note that the mean gas density $n_{\rm H} \simeq 9 \times 10^3$~cm$^{-3}$ derived from the dust SED analysis is consistent with the southern tip of the best fit area; the corresponding $\chi$ value is $\simeq 100$.

\cmark
The observed PACS fluxes toward the IC~1848 filaments allow an independent estimate of the UV radiation field strength \citep{kramer08,roccatagliata13}. We consider as an example position B near the head of the G1 filament, but away from the embedded IRAC point source. The PACS 70 and 160~\micron\ fluxes at this position are 212 and 598 MJy\,sr$^{-1}$, respectively (Fig.~3). The corresponding effective filter widths read from the filter curves in the PACS online documentation are $\sim 1.3 \times 10^{12}$ and $1.1 \times 10^{12}$ Hz, respectively. The UV field strength can then be estimated using eq. (2) of \cite{roccatagliata13} by adding the 70 and 160 $\mu$m PACS intensities \citep{schneider16}. The resulting value is $\chi = 74$ Habing units, corresponding to 96 Mathis units, consistent with the southern tip of the elongated best fit area in Figure~5.
\umark

\subsubsection{PDR models illuminated by a star}

The ISMDB models discussed above assume isotropic radiation and a normal incidence angle. To investigate the model results in more detail, we ran a series of offline PDR models with different geometries. The results are summarized in Table~4. The input parameters include the spectral type of the illuminating star (O5V is the closest spectral type considered in the Meudon code), distance $D$ (negative values imply front-side illumination and positive values back-side illumination), inclination angle $\theta$, $A_V$, and gas density $n_{\rm H}$.
\cmark
The UV radiation field of an OV5 star at a distance of 13 pc corresponds to $\chi \sim 35$ Mathis units (F. Le Petit, private comm.), lower than the values derived above. We note, however, that using a stellar spectral type as input parameter for the Meudon PDR code 
is different from using an ISRF scaling factor $\chi$ in the ISMDB analysis. $\chi$ is a scaling factor for an isotropic ISRF,  whereas the radiation field from a star is plane-parallel.
A plane-parallel (stellar) UV radiation field penetrates deeper into the cloud than an isotropic radiation field, resulting in higher intensities of FIR fine structure lines. The difference in the predicted line intensities can sometimes be significant.
\umark

Table~4 shows predicted fine structure line intensities and line ratios for several PDR models with different parameters computed using an off-line version of the Meudon PDR code (1.5.4).
Models 1027--1028 correspond to normal incidence at the closest possible (projected) distance from the illuminating star (highest possible illumination). Models 1029--1030 correspond to more physical geometries, with inclination angles $\theta = 30$ and 60$^\circ$, and the distance increased accordingly as the projected distance divided by $\sin \theta$. Models 1027 and 1029 are consistent with the observed line intensities of the  \CII\ and  \OI\ 63~$\mu$m lines, assuming front side illumination, although the \OI\ 63~$\mu$m/\CII\ line ratio is slightly lower than the observed values. However, the  \OI\ 145~$\mu$m/\CII\ line ratio for all models is an order of magnitude lower than the observed values, as suggested earlier by the ISMDB models.

While models with back-side illumination produce \OI\ 63/145~$\mu$m line ratios below unity, which is explained by foreground absorption of the optically thick \OI\ 63~$\mu$m line \citep{goldsmith21}, the predicted intensities of both lines are much too low compared to \CII. Therefore, we do not consider back-side illumination and optically thick \OI\ 63~$\mu$m emission as a viable solution for the IC1848 pillars.

Since under most conditions encountered here, the  \OI\ 145~$\mu$m line has a higher critical density than do the \CII\ and \OI\ 63~$\mu$m lines, we considered two additional models (1033--1034) with the same geometry as models 1029--1030, but with density increased by a factor of 5. While the intensity of the  \OI\ 145~$\mu$m line increases by a factor of 1.7, it is still well outside of the observed range. The \OI\ lines in IC1848 are weak and the S/N in the observed spectra is limited. Moreover, the \OI\ 145~$\mu$m line is close to a telluric ozone line, which affects the calibration of the data. However, we conclude that the apparent discrepancy between the intensities of the two \OI\ lines cannot be explained by instrumental effects.

Based on past experience \citep{joblin18, wu18}, the Meudon PDR code may underestimate the \CII\ line intensity in astronomical sources. This is explained by \CII\ contributions from gas in front or behind the PDR -- the model only reproduces the emission associated with the PDR. For oxygen lines, the plane-parallel geometry of the code can sometimes lead to problems, because photons from optically thick lines are artificially trapped in the model, whereas in reality they can escape in directions perpendicular to the line of sight. This line trapping effect is more important for the optically thick 63~$\mu$m line, with the optically thin 145\,$\mu$m line being little affected, and can artificially decrease the \OI\ 63/145~$\mu$m line ratio.
When comparing PDR models to observations, a scaling factor, $\kappa$, is sometime used to take into account that (a) several PDR layers along the line of sight can be present within the beam ($\kappa > 1$), (b) the PDR is inclined ($\kappa > 1$, an effect that we attempted to include directly in the analysis presented here), and (c) the PDR emission does not fill the beam ($\kappa < 1$). While the \CII\ emission in IC1848 is spatially extended, beam filling factors for different tracers may vary. Since we have only sparse single-point observations of the \OI\ lines with a limited S/N, we cannot ascertain the extent of the \OI\ emission, which is typically less extended than \CII\ given its higher critical density.

\cmark
\cite{rollig07} presented a comparison of numerical codes developed for modeling the physics and chemistry of UV illuminated regions. Their Figure~15 shows the surface brightness of the main fine-structure cooling lines computed using eight such models. With the exception of the Meijerink code, all models give the \OI\ 63/145~$\mu$m intensity ratio greater than unity (in main beam brightness temperature units). The \OI\ 63/145~$\mu$m ratio predicted by the Meudon PDR code is higher compared to other codes (e.g., a factor of $\sim 1.8$ higher that the KOSMA-$\tau$ model prediction). The particular model used for this comparison has a high UV illumination ($\chi = 10^5$) and the results may not be representative for all physical conditions. However, the differences among predictions of the various PDR models do not appear to be large enough to account for the discrepancy in the \OI\ 63/145~$\mu$m intensity ratio between the model predictions and our IC1848 observations.
\umark
Given the modeling and observational uncertainties (limited S/N, calibration uncertainty), a factor 2 -- 3 agreement with the observations is usually considered satisfactory. However, our observations suggest that the \OI\ 63/145~$\mu$m intensity ratio is a sensitive probe of the physical conditions (density or pressure, $\chi$) and chemistry in moderate $A_V$ and UV illumination regions such as IC1848.

\begin{figure}
   \centering
   \includegraphics[trim=2.5cm 2.75cm 2.25cm 4.5cm, clip=true, width=0.95\columnwidth]{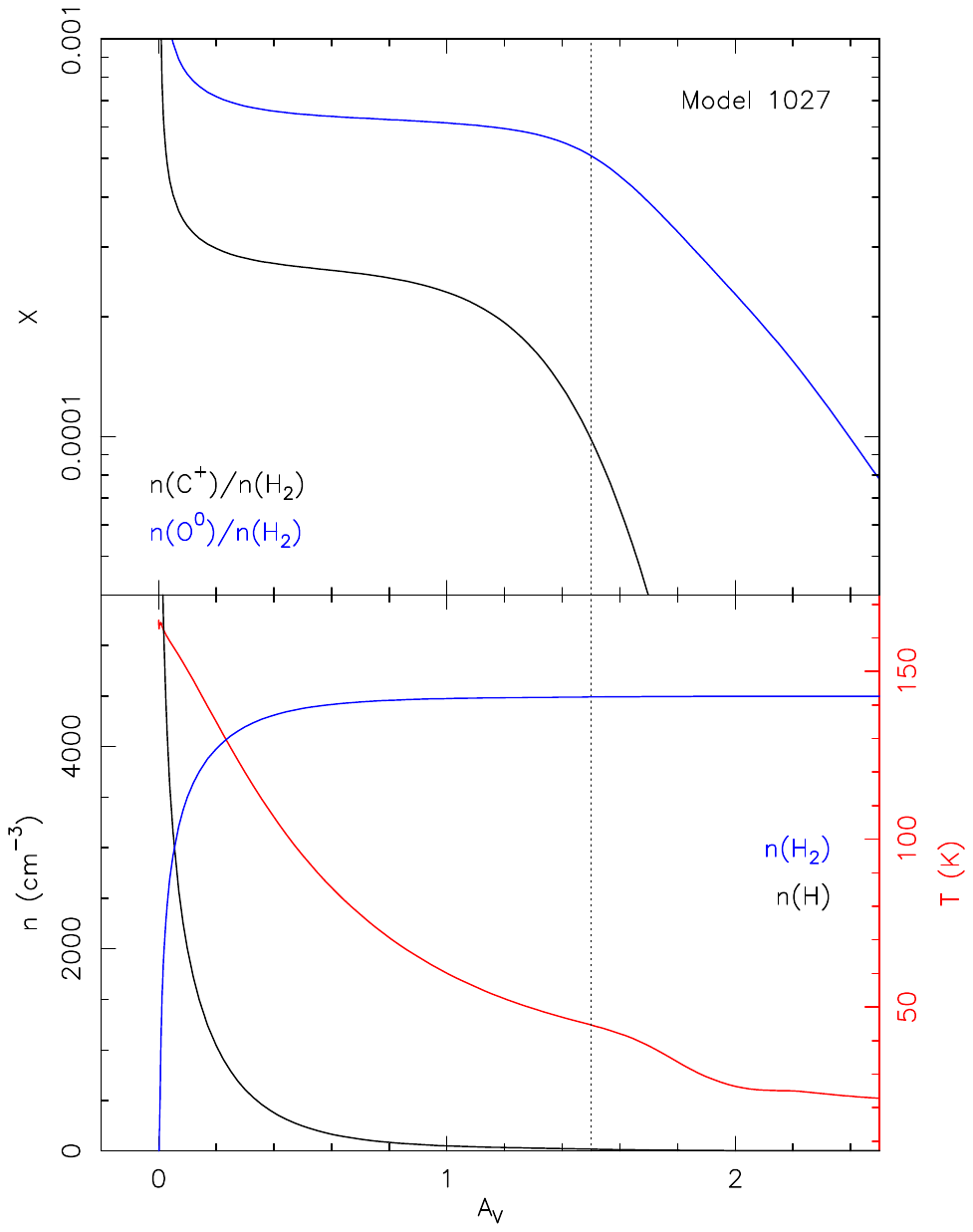} 
   \caption{Temperature, density, and abundance profiles in the outer PDR layers of Model 1027.} \label{fig:pdr}
 \end{figure}

Figure~\ref{fig:pdr} shows the gas temperature, density and abundance profiles near the cloud surface for Model 1027. PDR chemistry computations show that ionized carbon is only abundant in the outer $A_V \sim 1.5$ layer, transitioning to neutral carbon, CO, and other carbon-bearing species in deeper layers. Atomic oxygen extends somewhat deeper into the cloud up to $A_V \sim 2.5$, before being transformed into CO, water, and other oxygen-bearing species. The gas in the [C\,{\sc ii}] emitting region is molecular, with temperatures between 50 and 150 K.

\section{Summary and conclusions}
\label{sec:conclusions}

Using the upGREAT instrument on SOFIA, we have imaged the \CII\ 158~$\mu$m fine structure line in the bright-rimmed clouds G1--G3 in IC1848 and obtained pointed observations of the \OI\ 63 and 145~$\mu$m fine structure lines at several positions to characterize the morphology, velocity field, and the physical conditions in the region. The fine structure line intensities are analyzed using the Meudon PDR code. The main results of this study can be summarized as follows:

\begin{enumerate}

\item Velocity-resolved \CII\ spectra show evidence of a velocity shift at the head of the brightest G1 filament, possibly caused by radiation pressure from the impinging UV photons.

\item Archival \emph{Herschel} PACS and SPIRE images imply H$_2$ column densities in the range $10^{21} - 10^{22}$~cm$^{-2}$, corresponding to maximum visual extinction $A_V \simeq 10$. Assuming that the line-of-sight depth of the filaments is the same as their characteristic width in the plane of the sky, the average H$_2$ volume density is $\simeq 4500$~cm$^{-3}$.

\item The [C\,{\sc ii}] emission traces $\sim 17\%$ of the H$_2$ column density derived from the dust SED fits.

\item The \OI\ 63 and 145~$\mu$m emission is generally weak in this low-$\chi$ PDR. The 63~$\mu$m line is detected above 4~$\sigma$ level at 6 positions in G1 and 4 positions in G2, while the 145~$\mu$m line is detected only at 3 positions in G1. Both lines are detected at a high S/N in average spectra toward the head of the G1 filament (Figure 4), implying a low average \OI\ 63/145~\micron\ line intensity (in K\,kms$^{-1}$) ratio of $\sim 0.67$.

\item PDR models are unable to explain the observed line intensities of the two \OI\ fine structure lines in the IC1848 region. To obtain a good solution, the intensity of the \OI\ 145~$\mu$m line would have to be a factor of 3 lower compared to the observations. The \OI\ lines in IC1848 are weak and the S/N in the observed spectra is limited. However, the discrepancy between the intensities of the two \OI\ lines is not likely caused by instrumental/calibration effects.

\end{enumerate}

The apparent discrepancy between the model intensities of the two \OI\ lines and IC1848 observations may be due to the limited signal-to-noise ratio or trapping of the \OI\ 63~$\mu$m photons in a plane-parallel geometry model. Nevertheless, our observations suggest that the \OI\ 63/145~$\mu$m intensity ratio is a sensitive probe of the physical conditions (density or pressure, $\chi$) and chemistry in moderate $A_V$ and UV illumination regions such as IC1848. Antarctic balloons are the only current FIR facilities that would allow extending this work to other sources. However, some of the future FIR Astrophysics Probe concepts include high-resolution spectroscopic capabilities in this wavelength range.

\begin{acknowledgements}
  
Based on observations made with the NASA/DLR Stratospheric Observatory for Infrared Astronomy (SOFIA). SOFIA was jointly operated by the Universities Space Research Association, Inc. (USRA), under NASA contract NAS2-97001, and the Deutsches SOFIA Institut (DSI) under DLR contract 50 OK 0901 to the University of Stuttgart. GREAT was a development of the MPI für Radioastronomie and the KOSMA/Universität zu Köln, in cooperation with the DLR Institut für Optische Sensorsysteme, financed by the participating institutes, by the German Aerospace Center (DLR) under grants 50 OK 1102, 1103 and 1104, and within the Collaborative Research Centre 956, funded by the Deutsche Forschungsgemeinschaft (DFG). Part of this research was carried out at the Jet Propulsion Laboratory, California Institute of Technology, under a contract with the National Aeronautics and Space Administration (80NM0018D0004). D.C.L., P.F.G., and Y.S. acknowledge financial support from the National Aeronautics and Space Administration (NASA) Astrophysics Data Analysis Program (ADAP). We thank Franck Le Petit and William D. Langer
\cmark
and an anonymous referee
\umark
for helpful comments.

\end{acknowledgements}

\begin{appendix}\section{}

\begin{figure*}
   \centering
   \includegraphics[trim=2cm 2cm 8cm 1.5cm, clip=true, width=0.7\textwidth]{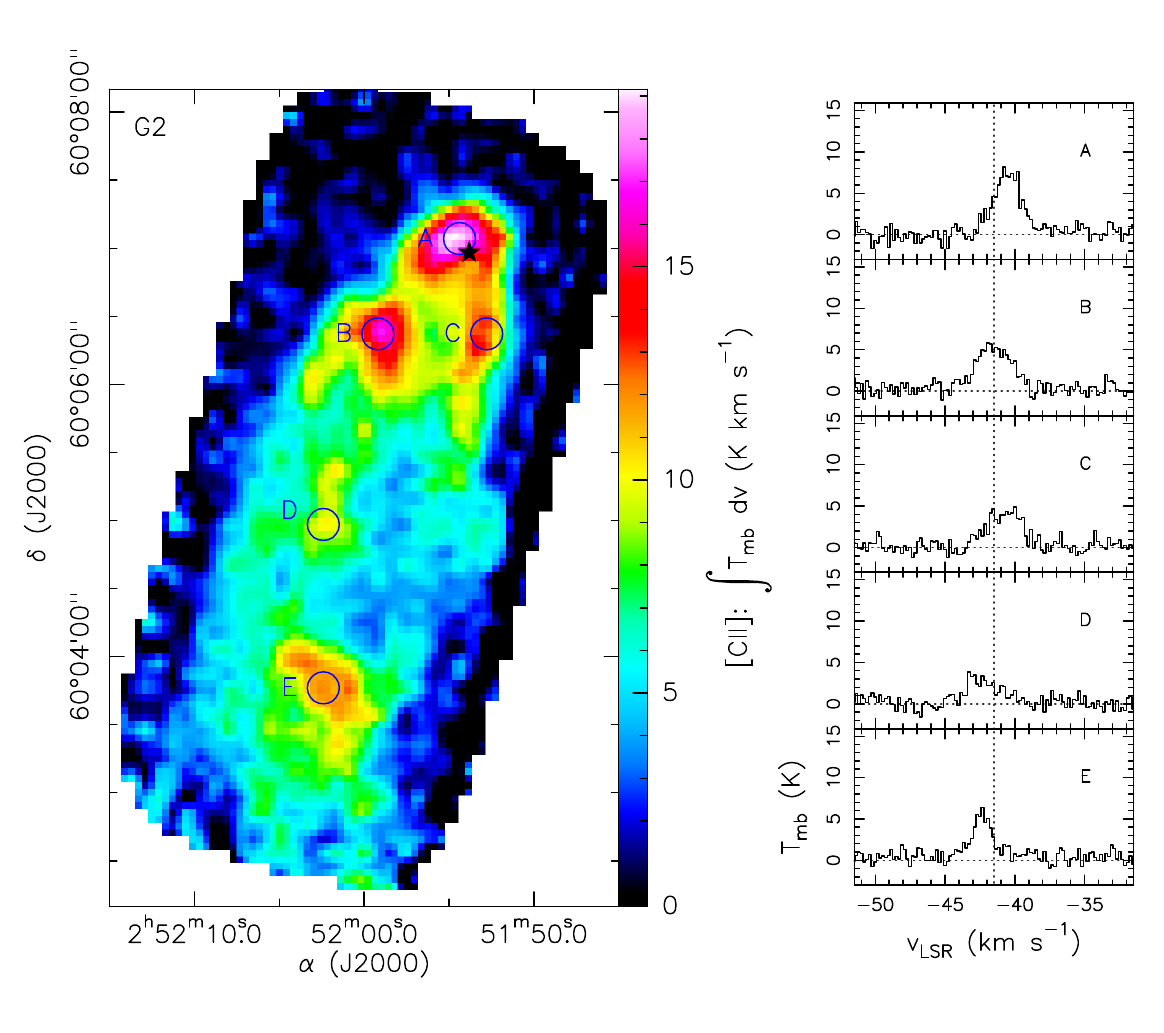} 
   \includegraphics[trim=2cm 2.cm 5.5cm 1.5cm, clip=true, width=0.7\textwidth]{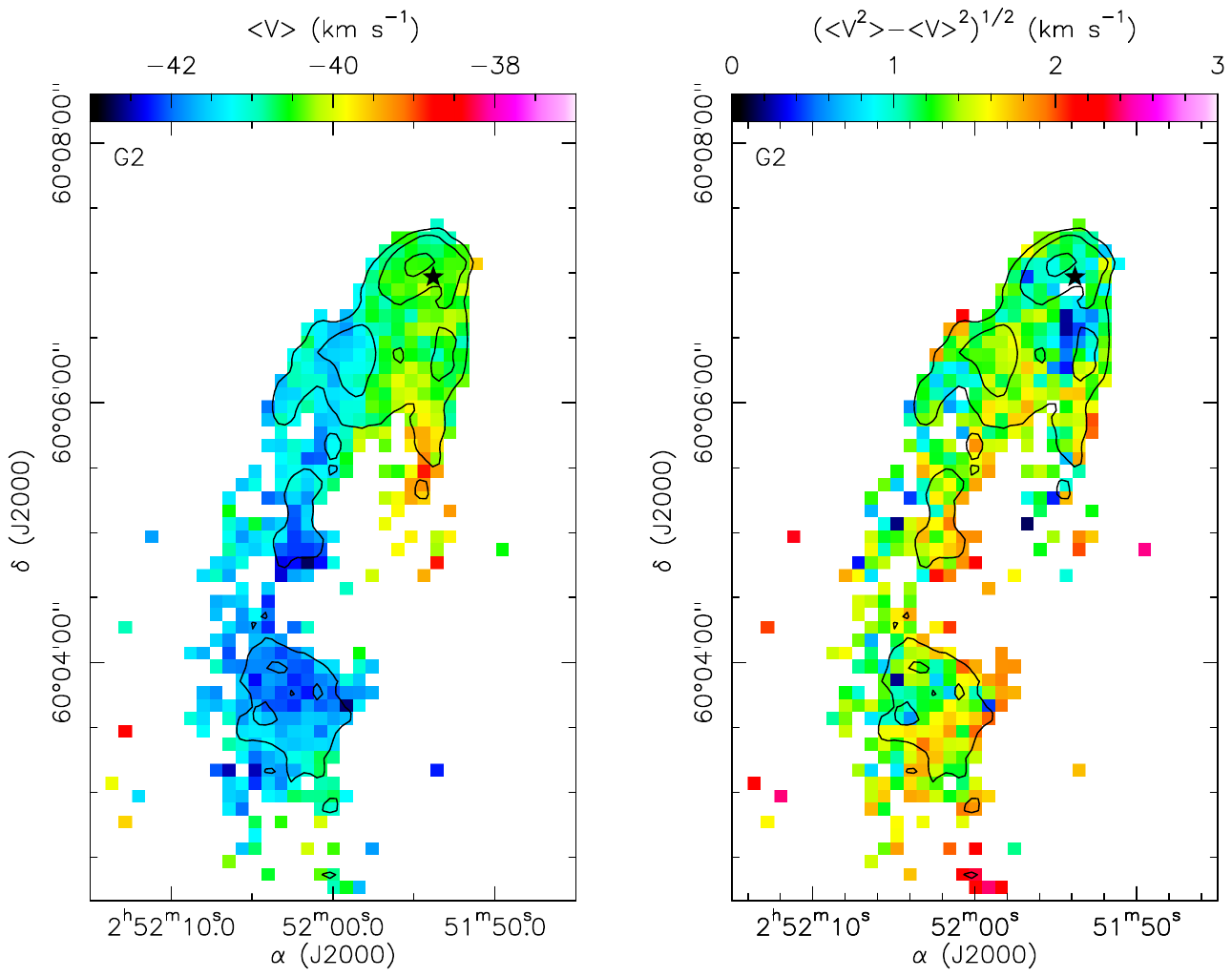} 

   \caption{(Top) Color image of the [C\,\,{\sc ii}] integrated line intensity in the --45 to --37 km\,s$^{-1}$ velocity range toward the G2 filament. The right panel shows spectra at selected positions labeled A--E in the image. Blue circles correspond to the FWHM SOFIA beam size at the [C\,\,{\sc ii} frequency. (Bottom) Maps of the [C\,\,{\sc ii}] line center velocity and velocity dispersion toward the G2 filament (color images, left and right panels, respectively). Only pixels with [C\,\,{\sc ii}] intensities above 3.5$\sigma$ (6.3 K\,km\,s$^{-1}$) are shown. Black contours show the integrated line intensity, with the same contour levels as in Figure~1. Black stars mark the location of the bright IRAC compact source.} \label{momentsg2}
\end{figure*}

\begin{figure*}
   \centering
   \includegraphics[trim=2cm 2cm 6cm 1.5cm, clip=true, width=0.7\textwidth]{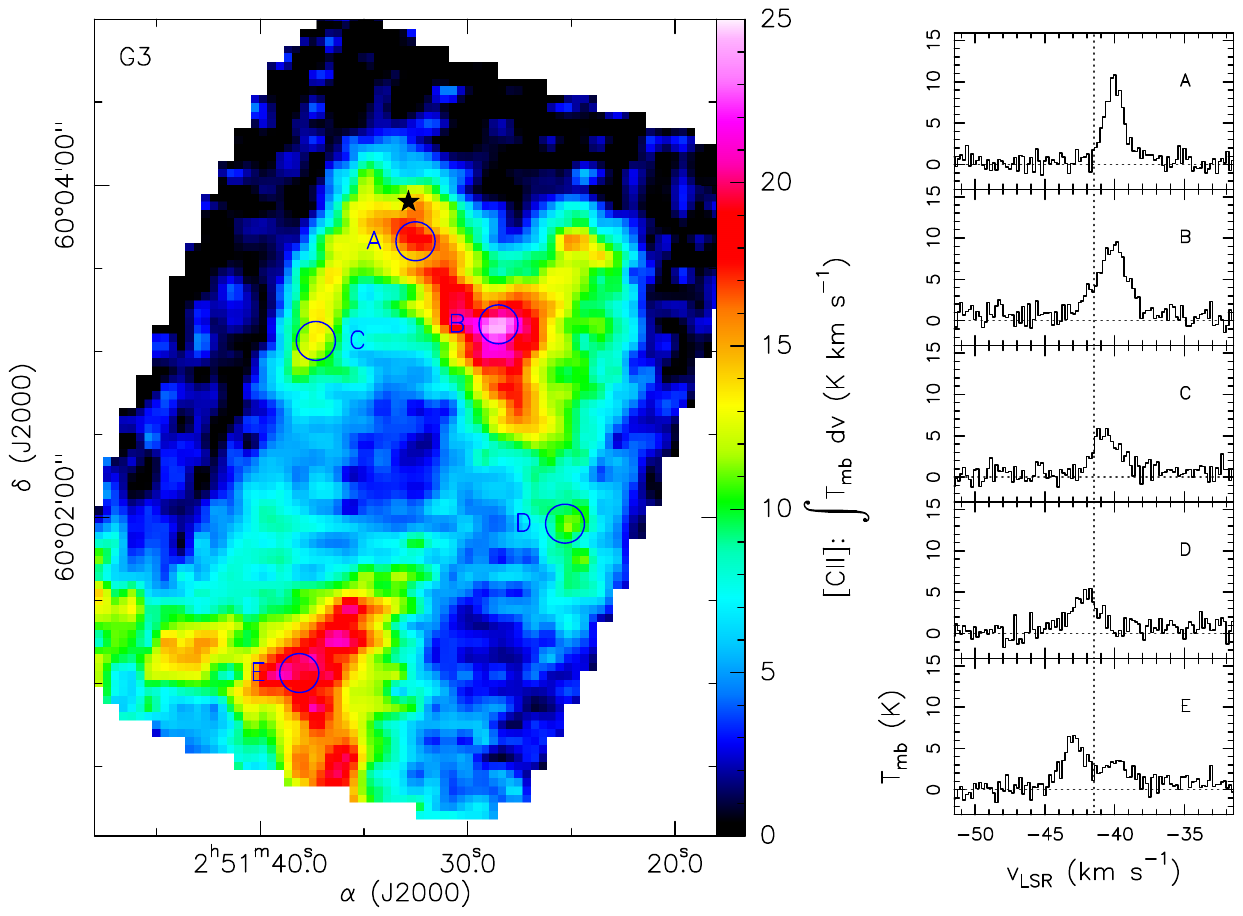} 
   \includegraphics[trim=1cm 2cm 5.5cm 1.5cm, clip=true, width=0.7\textwidth]{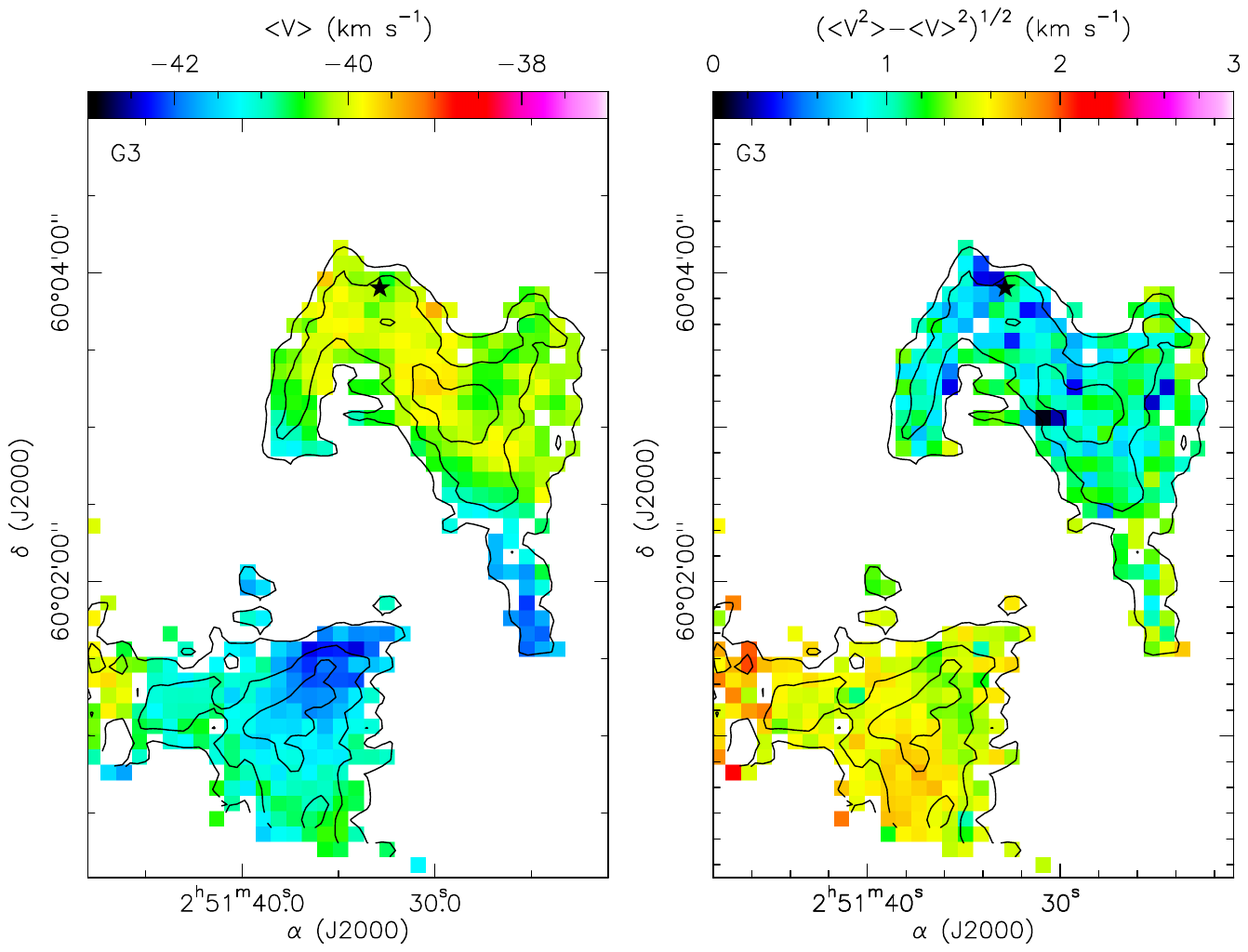} 

   \caption{(Top) Color image of the [C\,\,{\sc ii}] integrated line intensity in the --44 to --38 km\,s$^{-1}$ velocity range toward the G3 filament. The right panel shows spectra at selected positions labeled A--E in the image. Blue circles correspond to the SOFIA beam at the [C\,\,{\sc ii} frequency. (Bottom) Maps of the [C\,\,{\sc ii}] line center velocity and velocity dispersion toward the G2 filament (color images, left and right panels, respectively). Only pixels with [C\,\,{\sc ii}] intensities above 3.5$\sigma$ (8.4 K\,km\,s$^{-1}$) are shown. Black contours show the integrated line intensity, with the same contour levels as in Figure~1. Black stars mark the location of the bright IRAC compact source.} \label{momentsg3}
\end{figure*}

\begin{figure*}
   \centering
   \includegraphics[trim=2cm 3.75cm 6.4cm 3cm, clip=true, width=0.85\textwidth]{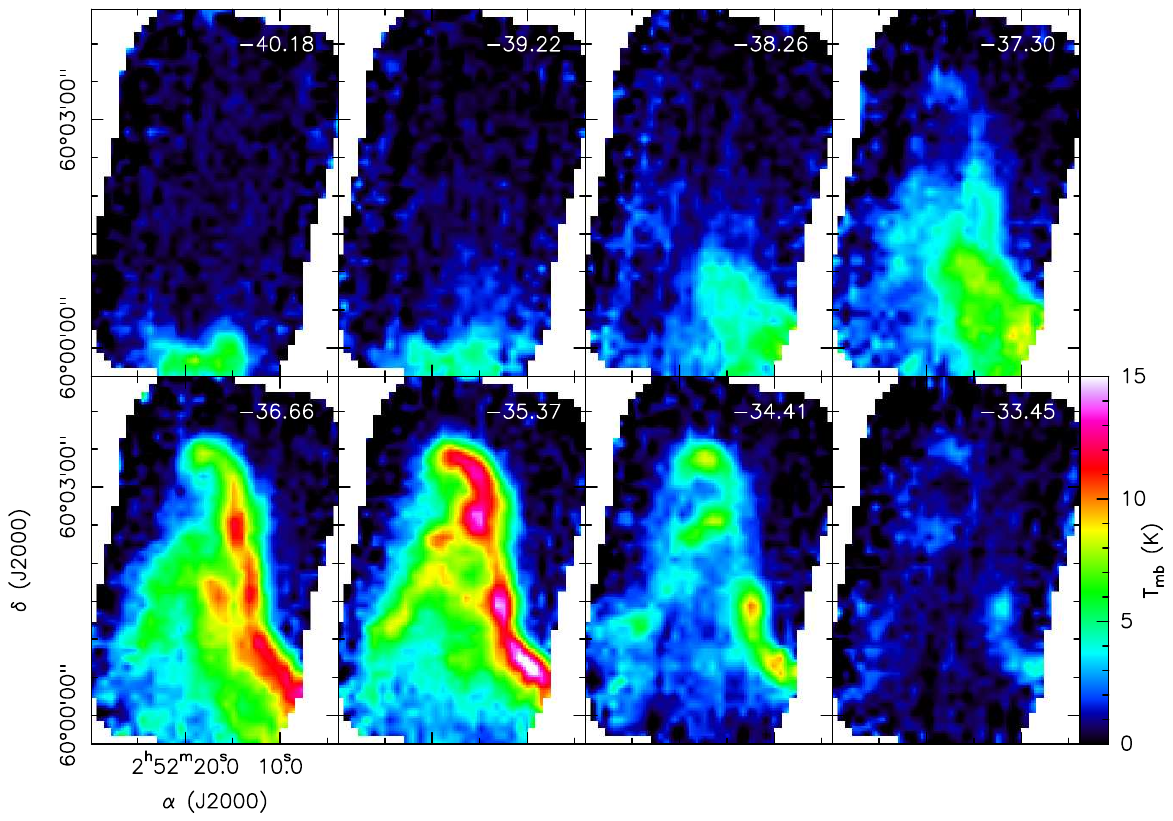} 

   \caption{Channel maps of the [C\,\,{\sc ii}] emission in the G1 filament. Center velocity of each channel is labeled in the upper-right corner.} \label{chang1}
 \end{figure*}

\begin{figure*}
   \centering
   \includegraphics[trim=2cm 3.75cm 8.5cm 3cm, clip=true, width=0.85\textwidth]{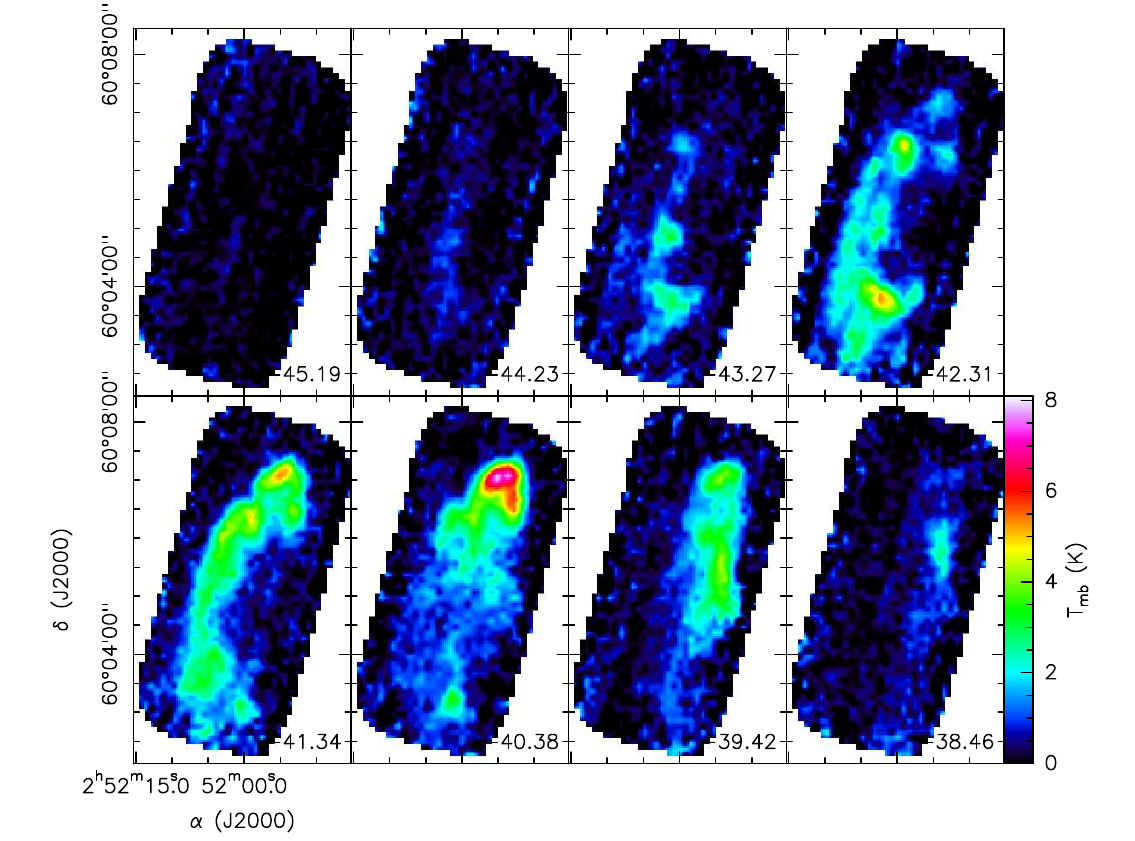} 

   \caption{Channel maps of the [C\,\,{\sc ii}] emission in the G2 filament. Center velocity of each channel is labeled in the lower-right corner.} \label{chang2}
 \end{figure*}

\begin{figure*}
   \centering
   \includegraphics[trim=2cm 4.5cm 5.5cm 3.5cm, clip=true, width=0.85\textwidth]{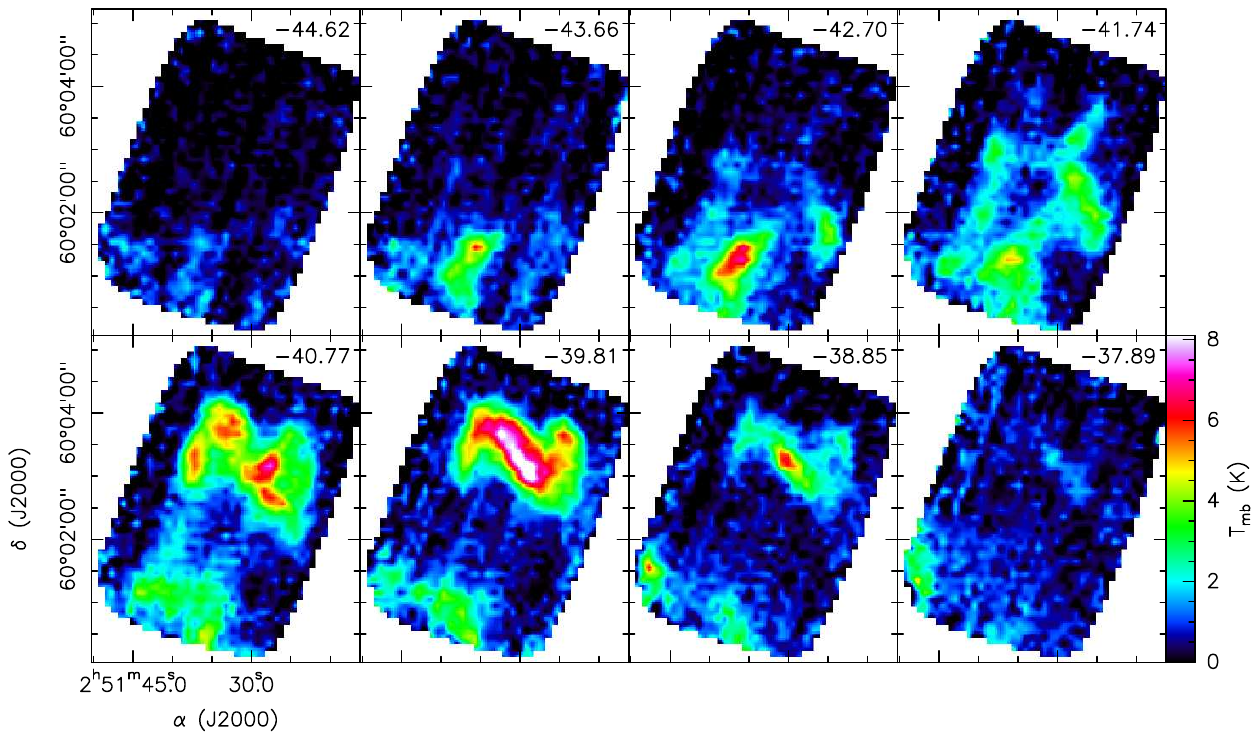} 

   \caption{Channel maps of the [C\,\,{\sc ii}] emission in the G3 filament. Center velocity of each channel is labeled in the upper-right corner.} \label{chang3}
\end{figure*}

\begin{figure*}
   \centering
   \includegraphics[trim=2.0cm 2.5cm 4cm 2.5cm, clip=true, width=0.8\textwidth]{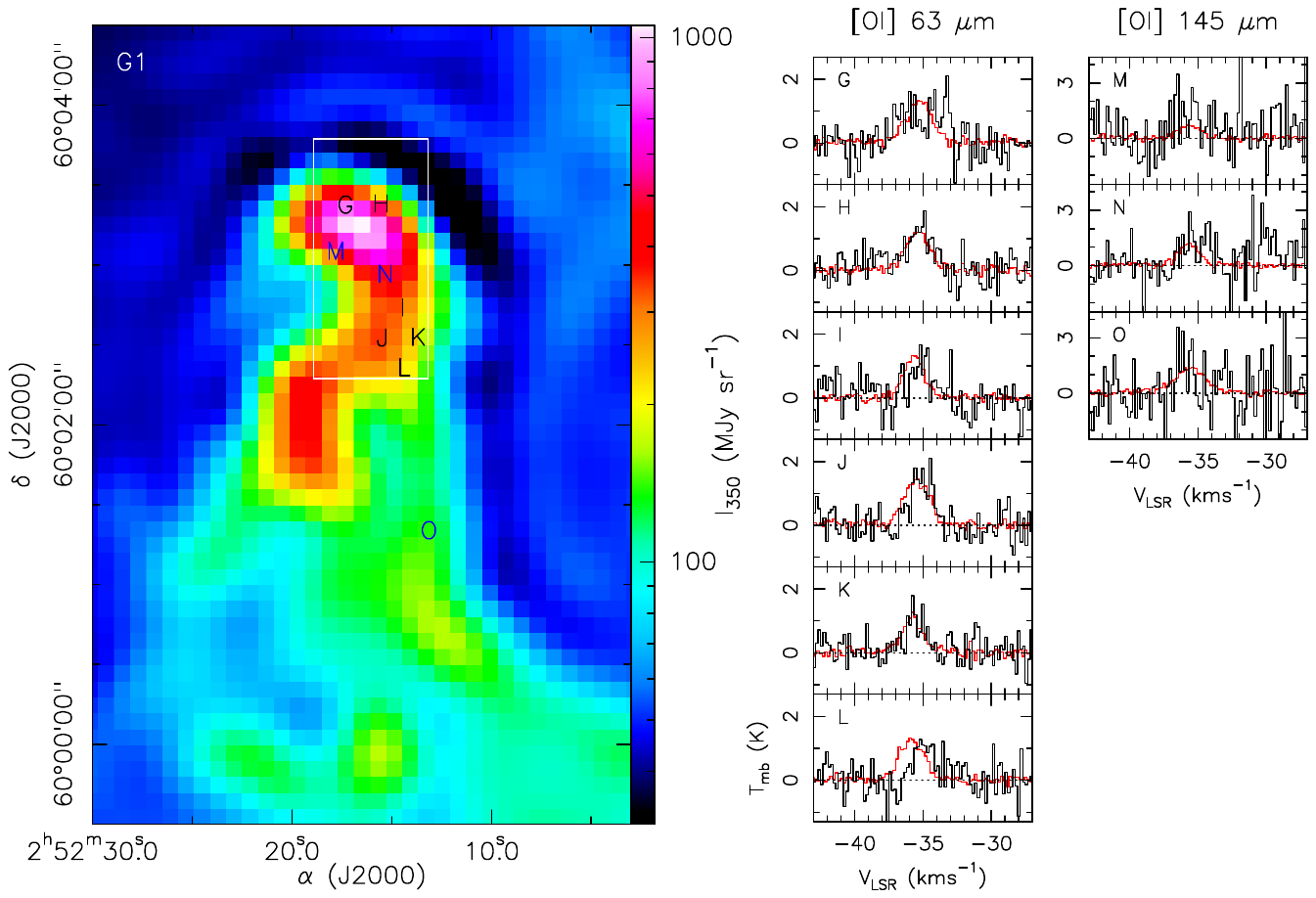} 
   \includegraphics[trim=2.0cm 2.5cm 4cm 2.5cm, clip=true, width=0.8\textwidth]{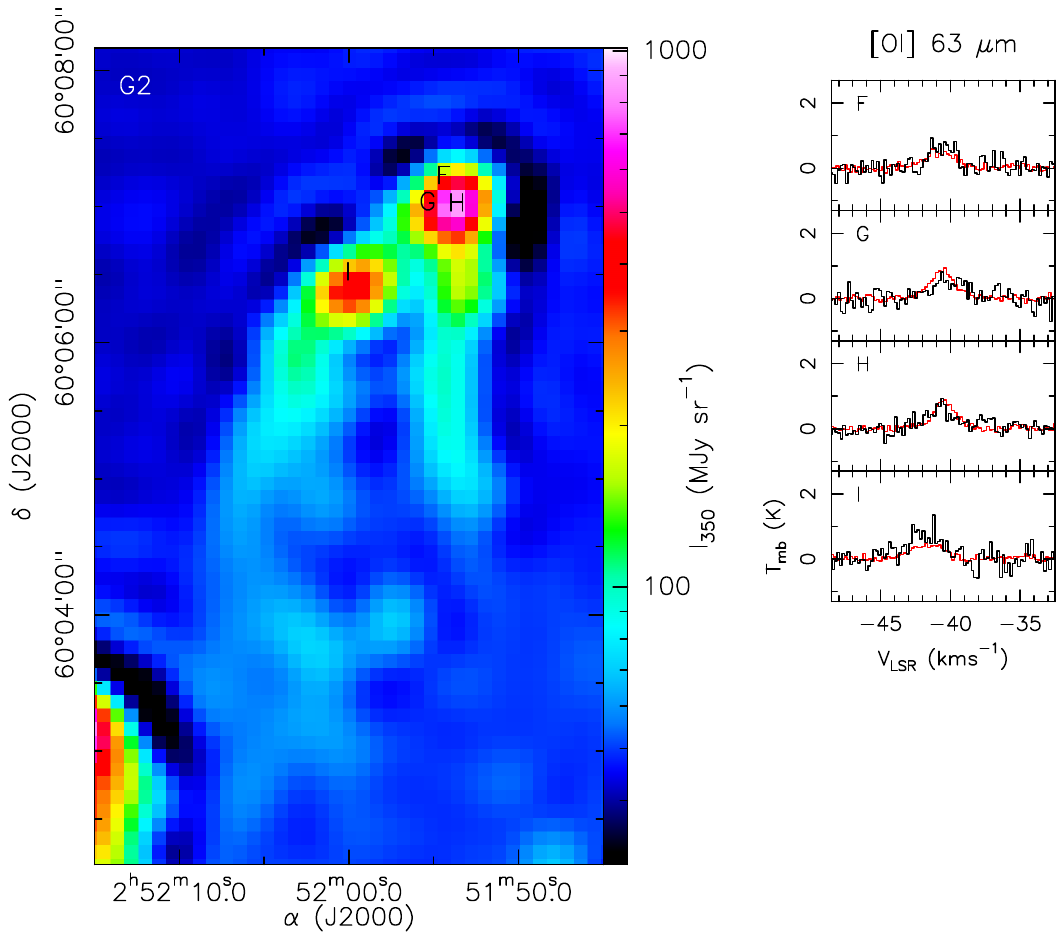} 
   \caption{[O\,{\sc i}] spectra (black histograms) toward positions G--O in the G1 filament (top) and positions F--G in the G2 filament (bottom). Red histograms show the corresponding [C\,{\sc ii}] spectra divided by 10. Positions at which the spectra are taken are marked on the SPIRE 350~$\mu$m images shown in the left panels.} \label{oxygen}
\end{figure*}

\begin{figure*}
   \centering 
   \includegraphics[trim=2.0cm 3.5cm 6cm 2.5cm, clip=true, width=0.8\textwidth]{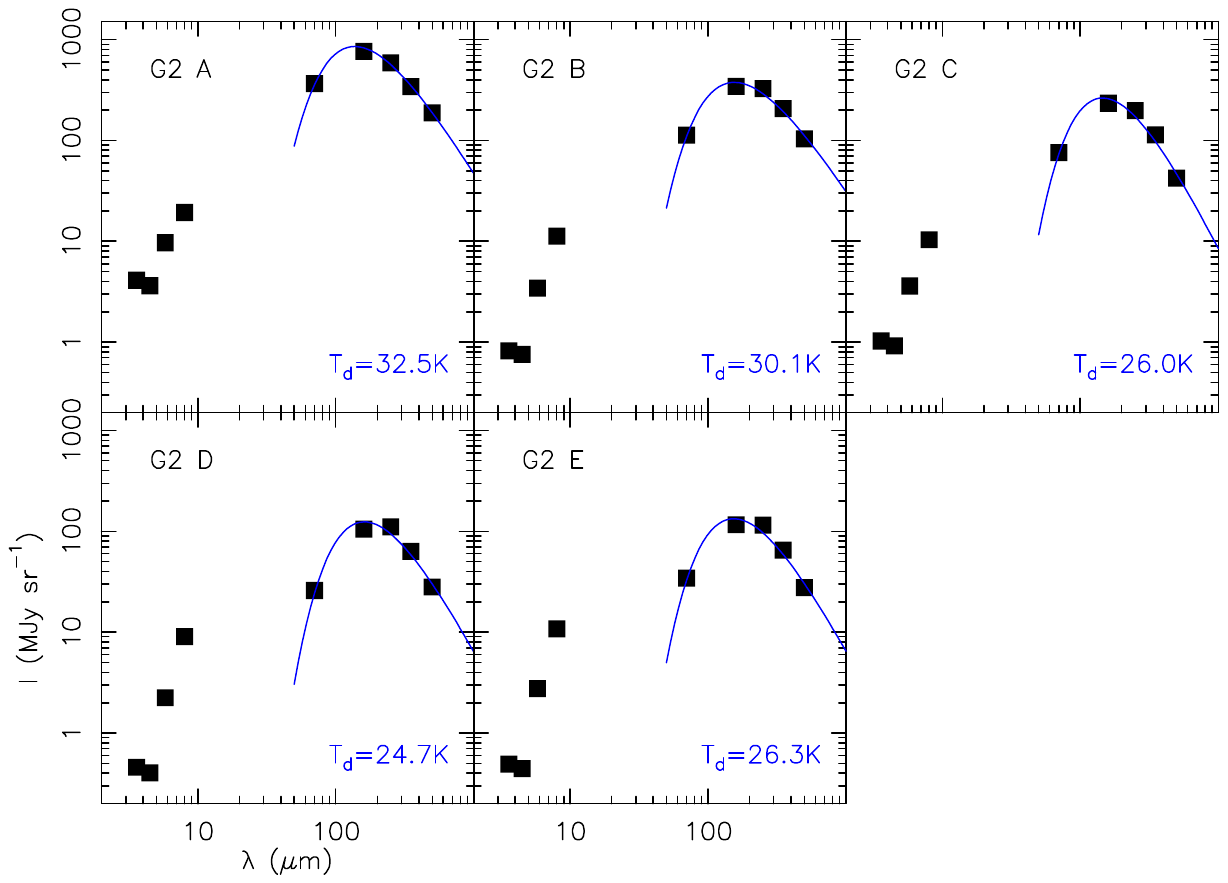} 
   \includegraphics[trim=2.0cm 3.5cm 6cm 2.5cm, clip=true, width=0.8\textwidth]{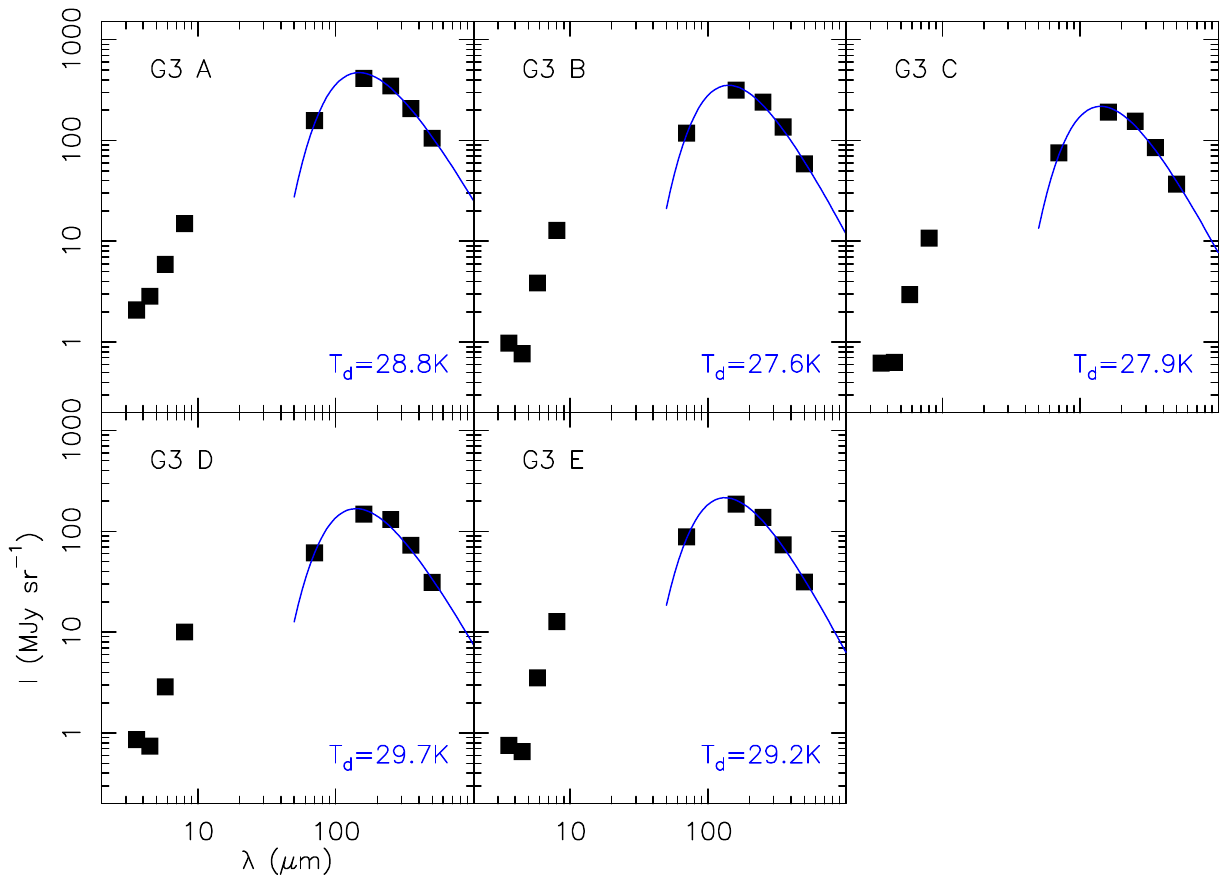} 
   \caption{Spectral energy distributions toward positions A--E in the G2 and G3 filaments, top and bottom respectively. Blue curves show the modified blackbody fits to the \emph{Herschel} PACS and SPIRE fluxes from which the dust temperatures and 350~$\mu$m optical depths used for the column density determination are derived.} \label{sedg2g3}
\end{figure*}

Figures~\ref{momentsg2} and \ref{momentsg3} show moment maps and [C\,{\sc ii}] spectra at selected positions toward the G2 and G3 filaments, respectively. Figures~\ref{chang1}--\ref{chang3} show the channel maps of the [C\,\,{\sc ii}] emission in the filaments G1--G3, respectively. Figure~\ref{oxygen} shows [O\,{\sc i}] spectra toward the G1 and G2 filaments. Figure~\ref{sedg2g3} shows the continuum SEDs at selected positions toward the G2 and G3 filaments.

\end{appendix}

\end{document}